\documentclass[reprint, showpacs, showkeys, amsmath, amssymb, floatfix, preprintnumbers, nofootinbib, superscriptaddress]{revtex4-1}
\pdfoutput=1
\usepackage{graphicx}% Include figure files
\usepackage{enumerate}% For non-numerical items
\usepackage{bm}% bold math
\usepackage{slashed}
\usepackage{comment}
\usepackage{bbold}
\usepackage{braket}
\usepackage{makecell}
\usepackage{amssymb}
\usepackage{epsfig}
\usepackage{epstopdf}
\newtheorem{lemma}{Lemma}

\newcommand{\be}{\begin{equation}}
\newcommand{\ee}{\end{equation}}
\newcommand\beq{\begin{eqnarray}}
\newcommand\eeq{\end{eqnarray}} 
\newcommand\eqn[1]{\label{eq:#1}} 
\newcommand\eq[1]{Eq.~(\ref{eq:#1})}

\newcommand\Eqs[2]{Eqs.~(\ref{eq:#1}-\ref{eq:#2})} 
\newcommand{\vev}[1]{\langle #1 \rangle}

\newcommand{\bfn}{{\mathbf n}}

\newcommand{\CA}{{\cal A}}

\newcommand{\CD}{{\cal D}}
\newcommand{\CE}{{\cal E}}
\newcommand{\CF}{{\cal F}}

\newcommand{\CP}{{\cal P}}

\newcommand{\gothn}{\mathfrak{n}}

\newcommand{\Tr}{{\rm Tr\,}}
\newcommand\half{{\textstyle{\frac{1}{2}}}}

\newcommand\expect[3]{\langle #1|#2|#3\rangle}

\newcommand{\mybar}[1]%
        {\kern 0.6pt\overline{\kern -0.6pt#1\kern -0.6pt}\kern 0.6pt}
\makeatletter
\renewcommand*\env@matrix[1][\arraystretch]{%
  \edef\arraystretch{#1}%
  \hskip -\arraycolsep
  \let\@ifnextchar\new@ifnextchar
  \array{*\c@MaxMatrixCols c}}
\makeatother

\begin{document}

\preprint{INT-PUB-16-032}

\title{A Chiral Solution to the Ginsparg-Wilson Equation}

\author{Dorota M. Grabowska}
 \email{grabow@uw.edu, dgrabowska@berkeley.edu}
\affiliation{Berkeley Center for Theoretical Physics, University of California, Berkeley, CA 94720}
\affiliation{Theoretical Physics Group, Lawrence Berkeley National Laboratory, Berkeley, CA 94720}

\author{David B. Kaplan}
\email{dbkaplan@uw.edu}
\affiliation{Institute for Nuclear Theory, Box 351550, Seattle, Washington 98195-1550, USA}
 
 \date{\today}
 
 \begin{abstract}
We present a chiral solution of the Ginsparg-Wilson equation.  This work is motivated by our recent proposal for nonperturbatively regulating chiral gauge theories, where five-dimensional domain wall fermions couple to a four-dimensional gauge field that is extended into the extra dimension as the solution to a gradient flow equation.  Mirror fermions at the far surface decouple from the gauge field as if they have form factors that become infinitely soft as the distance between the two surfaces is increased.  In the limit of an infinite extra dimension we derive an effective four-dimensional chiral overlap operator which is shown to obey the Ginsparg-Wilson equation, and which correctly reproduces a number of properties expected of chiral gauge theories in the continuum. 
 
% We present a chiral solution of the Ginsparg-Wilson equation.  This work is motivated by our recent 
% proposal for nonperturbatively regulating chiral gauge theories, where five-dimensional domain wall fermions couple to a four-dimensional gauge field that is extended into the extra dimension as the solution to a gradient flow equation.  Mirror fermions at the far surface decouple from the gauge field as if they have form factors that become infinitely soft as the distance between the two surfaces is increased.  At infinite extra dimension we can derive an effective four-dimensional chiral overlap operator which is shown to obey the Ginsparg-Wilson equation, and which  correctly reproduces a number of properties expected of a chiral gauge theory in the continuum. 

  \end{abstract}

 \maketitle
 \section{Introduction}
 Defining a nonperturbative regulator for chiral gauge theories has been a long-standing problem in quantum field theory.  While it may be that finding a regulator is just a technical issue, one should be open to the possibility that its resolution could entail new physics -- either in the form of new particles or  interactions, or through elucidation of some of the outstanding puzzles of the Standard Model, such as the strong CP problem. The difficulty in constructing a regulator comes down to defining a discretized version of the Euclidian fermion kinetic operator $\CD$ for Weyl fermions in a complex representation of the gauge group, where $\det \CD$  is the fermion contribution to the integration measure for  the gauge field path integral.  The naive target for the lattice theory is the kinetic operator  $\slashed{D} P_-$ ,  where  $D_\mu$ is the gauge covariant derivative and $P_\pm = (1\pm \gamma_5)/2$.  However, as has been discussed extensively in the literature, while the modulus of this determinant is given by the square root of the Dirac determinant, $|\det\CD |=\sqrt{\det\slashed{D}}$, its phase is not well-defined. The ambiguity in the phase arises because the operator $\CD$ maps negative chirality spinors into positive chirality spinors and therefore its eigenvalues cannot be uniquely defined. As the negative and positive chirality Hilbert spaces are independent, redefining each basis by an unrelated phase redefines the determinant by a phase which is an arbitrary functional of the gauge field (although most choices of phase could not result from a local fermion action). Furthermore, the phase of the determinant is only gauge-invariant when a theory has no gauge anomalies.

A definition of the Euclidian chiral determinant in the continuum was proposed in Ref.~\cite{AlvarezGaume:1983cs}, where the authors  introduced neutral spectators of opposite chirality, so that
that $\CD  =  \slashed{\partial} + i \slashed{A} P_-$, which in a chiral representation looks like
\beq
\CD = \begin{pmatrix} \ & D_\mu \sigma_\mu\\ \partial_\mu\bar\sigma_\mu & \end{pmatrix}  \ ,
\eqn{AG}\eeq  
with $\sigma_\mu = \{1,-i\vec\sigma\}$, $ \bar \sigma_\mu = \sigma_\mu^\dagger$.
 While this form of $\CD$ is not self-adjoint, it does have a well-defined eigenvalue problem and its  determinant can be uniquely determined~\cite{AlvarezGaume:1985di}.
 
This definition of $\CD$ cannot be directly implemented  on the lattice, 
as is evident when considering the global $U(1)$ chiral anomaly.  Chiral symmetry of the fermion action can be expressed by the equation $\{\CD,\gamma_5\}=0$, or equivalently (in the absence of exact zeromodes) as
\beq
\left\{\CD^{-1},\gamma_5\right\} = 0\ .
\eqn{cont5}
\eeq
However, in the continuum the path integral measure cannot be regulated in a way that preserves both gauge and chiral symmetries \cite{Fujikawa:1979ay}, which gives rise to the anomalous divergence of the axial current~\cite{Adler:1969gk, Bell:1969ab}
\beq
\partial_\mu j_\mu^5 = \frac{\alpha}{2\pi}\Tr F\widetilde F\ .
\eeq
 In contrast,  the path integration measure on the lattice is  defined in a way that is invariant under both gauge and chiral symmetries. Since it involves only a finite number of degrees of freedom, there are no anomalies and thus no anomalous divergence of the axial current.  The correct continuum limit with the axial anomaly  can therefore  only be attained if the lattice action is not invariant under chiral symmetry transformations.  In general such explicit symmetry breaking requires fine tuning to achieve a symmetry that is only broken anomalously in the continuum limit. Additionally at finite lattice spacing, important consequences of chiral symmetry, such as multiplicative mass renormalization, are usually lost.  Ginsparg and Wilson argued, however, that  by  modifying \eq{cont5} to read
\beq
\left\{\CD^{-1},\gamma_5\right\} = a \gamma_5\ ,
\eqn{latt5}
\eeq
where $a$ is the lattice spacing,\footnote{When zeromodes are present, one must use the equation of the operator itself, $\left\{\CD, \gamma_5\right\} = a \CD \gamma_5 \CD$.}  chiral symmetry would be broken in just the right way to reproduce the anomaly without fine tuning \cite{Ginsparg:1981bj}. It was subsequently shown that a solution to the Ginsparg-Wilson equation indeed gives rise to the correct anomaly, while at the same time ensuring an exact symmetry of the action at finite lattice spacing that enforces multiplicative mass renormalization and the absence of fine-tuning \cite{Hasenfratz:1998ri,Luscher:1998pqa}. In a chiral basis the general solution to \eq{latt5} is
\beq
\CD^{-1} = \begin{pmatrix} 0 & S_1\\ -S_2^\dagger & 0\end{pmatrix} + \frac{a}{2}\begin{pmatrix} 1 & 0\\  0 & 1\end{pmatrix}
\eqn{GW}
\eeq
where each block scales in size with the number of lattice sites, and $S_1$, $S_2$  can be independent operators; they are constrained by the desired continuum limit and locality, but not by \eq{latt5}. 

The Ginsparg-Wilson equation does not specify whether $\CD$ refers to fermions in a real (Dirac) or complex (chiral) representation of the gauge group.   Its solution in the Dirac case is given by the Narayanan-Neuberger   overlap operator \cite{Narayanan:1993sk,Narayanan:1994gw,Neuberger:1997fp,Neuberger:1998wv}. In this case  $S_1={S_2}$ and the effect of the diagonal term in $\CD^{-1}$ is very simple. The eigenvalues of the continuum Euclidian Dirac propagator $\slashed{D}^{-1}$ lie on the imaginary axis while the  eigenvalues  of $\CD^{-1}$  lie along a parallel line displaced from the imaginary axis by $a/2$; $\slashed{D}^{-1}$ has an infinite density of eigenvalues approaching the real axis while the lattice propagator $\CD^{-1}$ has a finite density, thanks to  the lattice cutoff.  The second term on the right side in \eq{GW} is responsible for the $a/2$ displacement and represents the explicit chiral symmetry breaking  that is required to reproduce the continuum anomaly.

For a  lattice regularization  of the target theory given in \eq{AG}  -- a chiral gauge theory with noninteracting mirror fermions -- we would expect that the solution \eq{GW} still pertains, but with $S_1\ne {S_2}$ and eigenvalues therefore no longer lying on a line parallel to the imaginary axis.  However, in this case the chiral symmetry violating part of the solution apparently requires either violating the gauge symmetry explicitly, or else allowing the mirror fermions to participate in the gauge interactions.  Either choice is a significant departure from the perturbative scheme. If gauge symmetry is explicitly broken, a path to restoring it in the continuum limit must be devised \cite{golterman2004s}; if the mirror fermions are gauged, one must understand how to  decouple them  in the continuum limit.  Both strategies have their theoretical challenges, and both have been pursued in the literature; we do not intend to review past work on the subject, but refer the reader to the review Ref.~\cite{Golterman:2000hr} as well as the  more recent papers Refs.~\cite{Luscher:2000hn,Wen:2013ppa,Giedt:2014pha,you2015interacting} and references therein.  

The focus of this paper is an alternative approach based on the proposal in Ref.~\cite{Grabowska:2015qpk}.  In this theory fermions of one chirality are surface modes on a five-dimensional slab coupling  to a gauge field $A$, while their mirror partners of the opposite chirality are modes on the opposite surface coupling to a different gauge field $A_\star$.   The two gauge fields are related by a gauge-covariant flow equation, where the  field $A$ on one surface  flows to $A_\star$ on the other. In the limit of infinite extra dimension we find a solution to the Ginsparg-Wilson equation of the form \eq{GW} with the continuum limit
\beq
\lim_{a\to 0} \hat\CD_\chi 
=\begin{pmatrix}0&  \sigma_\mu D_\mu(A) \\ \bar\sigma_\mu D_\mu(A_\star) &0\end{pmatrix}\ .
\eqn{AGchi}\eeq
Since we only consider gauge-covariant flow equations, the gauge fields $A$ and $A_\star$ transform identically under gauge transformations and the 
  diagonal entries of $\CD^{-1}$ at nonzero lattice spacing do not violate gauge invariance.   In the limit of infinite extra dimension, $A_\star$ is the fixed point of the flow equation given the initial data $A$.  We will be interested in two possible scenarios: one where $A_\star$ is the classical multi-instanton solution with winding number equal to that of $A$, and the other where $A_\star$ is pure gauge, the latter being a possible fixed point for a gauge covariant gradient flow equation on the lattice.  In either case, with all dynamical degrees of freedom damped out of $A_\star$, one might expect the mirror fermions to entirely decouple in the continuum and infinite volume limits, effectively realizing the continuum construction in \eq{AG}.
 
  In the next section we review the proposal of Ref.~\cite{Grabowska:2015qpk} for five-dimensional  fermions coupled to a four-dimensional gauge field, extended into the extra dimension via gradient flow. We then review the  technology developed by  Narayanan and Neuberger to construct the effective overlap fermion operator  for vector-like gauge theories from domain wall fermions with infinite extra dimension~\cite{Narayanan:1993sk,Narayanan:1994gw,Neuberger:1997fp,Neuberger:1998wv}. By applying their reasoning to the theory of Ref.~\cite{Grabowska:2015qpk} we attain the main result of this paper.   After discussing the behavior of the chiral overlap operator for gauge fields with nontrivial topology we  suggest a simulation to test key ideas presented here.\footnote{Preliminary versions of this work were presented at the 34th International Symposium on Lattice Field Theory in Southampton, UK, July 24-30, 2016  \cite{Grabowska:Lattice, Kaplan:Lattice}.}
  
\section{Domain wall fermions for chiral gauge theories}

Domain wall fermions can be formulated as Dirac fermions in five Euclidean dimensions with masses that depend on the extra dimension. Specifically, consider the coordinate of the extra dimension to be $s\in [-L, L]$, with periodic boundary conditions for a Dirac fermion field which has a positive mass on half the space and a negative mass on the other half  \cite{Kaplan:1992bt,Kaplan:2009yg,Grabowska:2015qpk}. The spectrum contains a light boundstate at each of the two mass defects that behave as a four-dimensional Dirac fermion with a mass which vanishes exponentially fast in the $L\to\infty$ limit; the two boundstates become positive and negative chirality eigenstates respectively in that limit.   The domain wall fermion construction provides a  solution to the problem of realizing chiral symmetry correctly for lattice  fermions in a  vector-like representation of the gauge group:  (i) the chiral anomaly is correctly realized via the Callan-Harvey effect, where a Chern-Simons operator is generated by integrating out the massive bulk fermions~\cite{Callan:1984sa},  and (ii) any small mass term introduced for the light modes can only be multiplicatively renormalized due to the vanishing wavefunction overlap between the negative and positive chirality fermion modes in the absence of such a mass term.  Furthermore  the number and chiralities of the light surface modes in the spectrum is a topological invariant of the bulk fermion dispersion relation in momentum space, as shown in Ref.~\cite{Golterman:1992ub}, and is not simply given by the number of fields in the five-dimensional theory.

An important feature discovered in Refs. \cite{Jansen:1992tw,Golterman:1992ub} is that in a regulated theory the contributions to the Chern-Simons current can trivially vanish in half of the bulk, making that region a true insulator.  Thus one can  take the fermion mass to be infinite on the $s\in(-L,0)$ half space, essentially excising that part of the space,  and instead only consider the half-space $s\in [0,L]$, with fermion zeromodes confined to the surfaces of the slab. This gives rise to Shamir's formulation of domain wall fermions \cite{Shamir:1993zy,Furman:1994ky} which serves as the foundation for practical simulations of lattice QCD. 

From the beginning, the physical separation of chiral fermion modes in the extra dimension has been seen as a promising starting point for the construction of chiral gauge theories   \cite{Kaplan:1992bt}.  Multiple flavors of fermions in different representations of the gauge group can be trivially incorporated in the construction, with either chirality   localized at a specific surface.  Of particular interest are systems where the coefficient of the bulk Chern-Simons operator in the effective theory vanishes due to cancellations between contributions from the different flavors of fermions, eliminating charge exchange between the two surfaces.  Such a cancellation is equivalent to the group theoretical statement that the chiral fermions at each surface are independently in a representation that is free from gauge anomalies, and hence each is in its own right a candidate for a healthy four-dimensional gauge theory. Constructions of anomaly cancellation models were examined in the early 1990s  \cite{Jansen:1992yj,Kaplan:1992bt,Kaplan:1992sg}, the most trivial consisting of two five-dimensional (or three-dimensional) fermions in the same gauge representation but with opposite signs for their masses, rendering the theory $P$- and $T$-invariant, with the zeromode spectrum at each surface consisting of a massless Dirac fermion. Quantum field theories exhibiting less trivial anomaly cancellation were also studied, such as the  3-4-5 $U(1)$ gauge theory in two dimensions~\cite{Jansen:1992tw}. Models where the surface modes are Majorana fermions  were also constructed for the purpose of simulating supersymmetry, where the massless Majorana fermions serve as gauginos \cite{Neuberger:1997bg,Kaplan:1999jn}.

These phenomena have direct analogues in condensed matter physics.  The Chern-Simons operator with its quantized coefficient and  bulk current flow describes the integer quantum Hall effect.  The most basic model  which exhibits anomaly cancelation, giving rise to a massless Dirac fermion at each defect,  is the same model rediscovered over a decade later in condensed matter systems by Kane and Mele \cite{kane2005quantum}, and the anomalous flow of global chiral charge between the two surfaces has been dubbed the ``Quantum Spin Hall Effect". Such materials  are referred to as topological insulators because of the topological stability of the zeromodes as shown in Ref.~\cite{Golterman:1992ub}.  Models with Majorana surface modes were rediscovered in the context of quantum computation in Ref.~\cite{kitaev2001unpaired}.  We are not aware of condensed matter analogues of the  less trivial anomaly cancellation models, however, such as the 3-4-5 model of Ref.~\cite{Jansen:1992yj}.

While it is easy to construct models with  chiral fermion representations for which gauge anomalies cancel at each surface of the extra dimension theory, it is a challenge to eliminate gauge couplings between the fermions at one surface from their mirror partners at the other.  This difficulty is closely related to the problem discussed  in the previous section:  the domain wall construction correctly reproduces the chiral $U(1)$ anomaly by coupling the chiral modes at the two surfaces to each other via the bulk fermions, and gauge invariance then requires that they both couple to the same gauge fields.  This automatically gives rise to a vector-like gauge theory in the continuum unless one   either constructs a mechanism to gap the mirror fermions in a gauge-invariant way, localizes the gauge fields in the extra dimension in the region near one of the surfaces, or resorts to explicit breaking of gauge invariance.  Numerous attempts to gap the mirror fermions have failed (see, for example Ref.~\cite{Giedt:2014pha}) as have attempts to eliminate mirror fermions by localizing the gauge fields  (reviewed in Ref.~\cite{Golterman:2000hr}).

An alternative was proposed in Ref.~\cite{Grabowska:2015qpk}.  A four-dimensional gauge field with an s-dependent profile, $\CA_\mu(x, s)$, is defined throughout the five-dimensional space $s\in [-L,L]$ as the solution to a gradient flow equation with periodic boundary conditions; the field is even in $s$.\footnote{The field called $\CA(x, s)$ here was referred to as $\bar A(x,s)$ in Ref.~\cite{Grabowska:2015qpk}.} As the gradient flow equation is a first order differential equation in $s$, the field throughout the bulk is determined by the value of the field at $s=0$, which is also the four-dimensional gauge field $A(x)$  appearing as the integration variable of the path integral. For positive $s$ the continuum version of the flow equation advocated is the conventional one discussed in the literature \cite{atiyah1983yang,Narayanan:2006rf,Luscher:2010iy}, 
\beq
\partial_\tau \CA_\mu =\text{sgn}(\tau) D_\nu \CF_{\mu\nu}\ ,\qquad  \CA_\mu(x,0)=A_\mu(x)
\eqn{flow}\eeq
where $D_\nu$ and $\CF_{\mu\nu}$ are constructed from $\CA_\mu(x,s)$, with Lorentz indices running over $\mu = 1,\ldots,4$. Here the ``flow time" $\tau(s)$ is taken to be an odd  monotonic  function of the coordinate $s$. The only reason we do not take $\tau(s)=s$ is because we wish to consider the case of $\CA$ changing abruptly with $s$ in between the two surfaces at $s=0$ and $s=\pm L$.

The effects of gradient flow are simply explained with the example of a two-dimensional $U(1)$ gauge field.  In this case the gauge field decomposes into a physical degree of freedom $\lambda$ and a gauge degree of freedom $\omega$,
\beq
\CA_\mu(x,s) = \epsilon_{\mu\nu}\partial_\mu \lambda(x,s) + \partial_\mu\omega(x,s) \, ,
\eeq
where $\omega$ shifts under a gauge transformation while $\lambda$ is invariant. The gradient flow equation \eq{flow} for the Fourier transforms of $\lambda$, $\omega$ takes the form
\beq
\partial_s \tilde \lambda(p,s) = -\text{sgn}(s)\frac{p^2 \tilde \lambda(p,s)}{\Lambda} \qquad \partial_s \tilde \omega(p,s) = 0 \, ,
\eeq
where we have set $\tau(s)=\Lambda s$ and $\Lambda$ is some mass scale.  The solutions are
\beq
\tilde \lambda(p,s)  = e^{-p^2 |s|/\Lambda}\tilde\lambda(p,0) \qquad \tilde \omega(p,s) =  \tilde\omega(p,0)  \, ,
\eeq
where $\tilde\lambda(p,0)$ and $ \tilde\omega(p,0)$ are the boundary data for the gauge field at  $s=0$. We see that gradient flow causes the physical field $\tilde \lambda$ to vanish exponentially fast in $s$ with an exponent proportional to $p^2$, while the gauge degree of freedom $\omega$ is unaffected.  A fermion which interacts with $\CA_\mu(x,s)$ therefore appears to be a charged particle, but one with a gaussian form factor, behaving as a particle of size  $\sqrt{s/\Lambda}$. Thus the fermion zeromodes localized at $s=0$ look like ordinary particles, while those localized at $s=L$ effectively  have  size $\ell = \sqrt{L/\Lambda}$ and are dubbed ``fluff"; their   size grows as $L$ increases and in the limit $L\to\infty$ they become incapable of exchanging momenta with other fermions via gauge boson exchange.  In that limit the physical gauge field $\lambda$ decouples from the mirror fermions and they only see the gauge degree of freedom $\omega$; gradient flow acts as a projection operator on the initial gauge field, eliminating all physical degrees of freedom. This behavior, that gradient flow only smooths out the physical degrees of freedom, persists when looking at nonabelian groups. Therefore, for any gauge group the chiral zeromodes at $s=0$ interact mainly with  gauge field $A(x)$, while the mirror fermions at $s=L$ interact with $A_\star(x) $, where
\beq
A_\star(x) = \CA(x,L)\ .
\eeq
In the above example, the gauge covariance of the flow equation \eq{flow} under $s$-independent gauge transformations is reflected in the $s$-independence of the solution for $\omega$. This guarantees that the fields $A$ and $A_\star$ at the two domain walls transform identically under gauge transformations, as do the zeromodes residing there. Since the fermion fields have five-dimensional dynamics, this makes our application of gradient flow quite different from preceding applications, having physical consequences rather than simply being a regularization scheme to smooth out operators.
 
One feature of gradient flow  is oversimplified in the above example.   No continuous flow equation can change the topology of the gauge field and therefore  a nonabelian gauge field $A$ in four dimensions characterized by a winding number $\nu$ can only flow to a gauge field $A_\star$ with winding number $\nu_\star=\nu$. Thus for $\nu\ne 0$  $A_\star$ cannot be pure gauge.  More generally, any gauge field $A$ will flow at large $L$ toward an attractive fixed point of the flow equation, which for \eq{flow} implies a stable solution to the Euclidian equations of motion. In each topological sector these fixed points include at least all of the classical multi-instanton (or anti-instanton) solutions.  It seems plausible that these are in fact the only attractive fixed points, and  in particular that there are no attractive fixed points containing an instanton-anti-instanton pair.   As discussed in Ref.~\cite{Grabowska:2015qpk}, having topological correlations between $A$ and $A_\star$ will induce correlations (e.g. interactions) between the zeromodes at the two surfaces, even at infinite domain wall separation.  However these correlations will be neither local nor extensive, and it is unclear whether they survive the infinite volume limit.  It was left as an open question there whether these unusual topological properties of the theory could be exploited to solve the strong CP problem in QCD, a question recently revisited in more detail in \cite{Okumura:2016dsr}.

%%%%%%%%%%
\begin{figure*}[t]
\includegraphics[width=15cm]{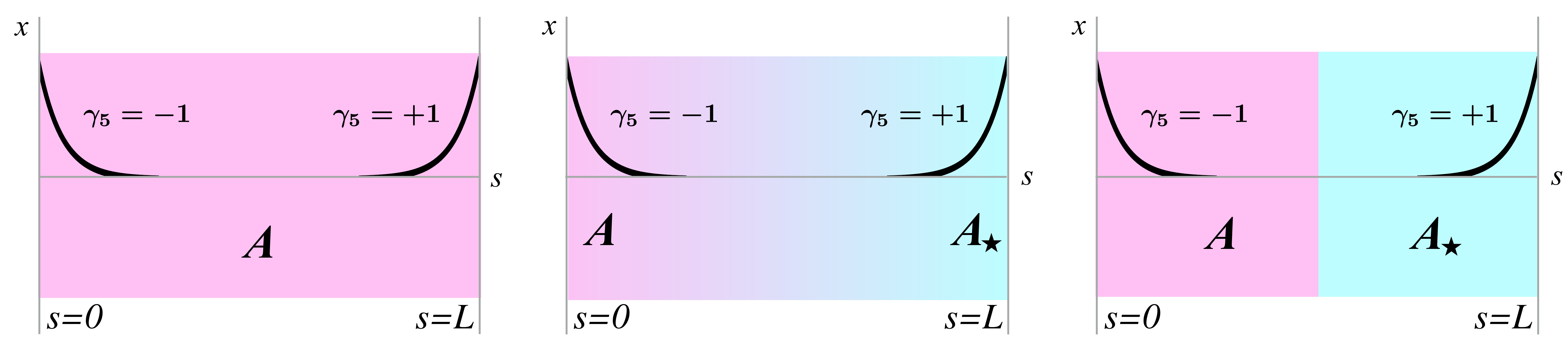}
\caption{  \it Left: the conventional domain wall fermion construction, with   negative and positive chiral modes bound to the $s=0$ and $s=L$ surfaces of the extra dimension respectively; the gauge field $A$ is constant over the extra dimension.  Center: the chiral construction where the gauge field in the extra dimension is a solution of the gradient flow equation, flowing from initial value $A$ on the left to the corresponding fixed point value $A_\star$ on the right. Right: Flow which makes an abrupt transition from $A$ to $A_\star$, a case that is more easily treated analytically. When assuming this scenario we place hats on our lattice operators, such as $\hat \CD_\chi$.  }
\label{fig:vcoverlap}
\end{figure*}
%%%%%%%%%%%%
 
The order of limits taken in  Ref.~\cite{Grabowska:2015qpk} -- where the continuum limit was taken first, and the resulting theory considered at large but finite $L$ -- plays a crucial role in the above discussion  about gauge field topology.  Of course, finite $L$ is required if the five-dimensional theory is to be numerically simulated.  However, the mirror fermions in such a theory   couple to gauge fields with gaussian form factors $\exp(-p^2L/\Lambda)$ and while such form factors may be very small in Euclidian space, they have no sensible analytic continuation to Minkowski spacetime: $p^2$, the square of the gauge boson momentum transfer, is not positive definite in Minkowski spacetime  and so  such form factors can diverge  for large $L$ or large negative $p^2$.  For this reason we consider in this paper the opposite order of limits, taking the infinite $L$ limit at finite lattice spacing first, before taking the continuum limit.  As Narayanan and Neuberger showed, there exists at finite lattice spacing a relatively simple four-dimensional description of vector-like domain wall fermions in the $L\to\infty$ limit in terms of  the overlap operator \cite{Narayanan:1993sk,Narayanan:1994gw, Neuberger:1997bg,Neuberger:1997fp,Neuberger:1998wv}.   Therefore in this paper we apply  their analysis to the chiral domain wall theory with gradient flow, seeking a purely four-dimensional description of the five-dimensional lattice theory in the $L\to\infty$ limit.  

One immediate consequence of working at finite lattice spacing is that the flow equation \eq{flow} must be replaced by a discrete version, such as the discretized Wilson flow equations for link variables discussed in Ref.~\cite{Luscher:2010iy}. It is believed that Wilson flow has no nontrivial attractive fixed points~\cite{luscher1982, Teper:1985rb, Teper:1985gi, Teper:1985ek}, in which case the discrete version of the $A_\star$ gauge field experienced by the mirror fermions can be pure gauge with $\nu_\star=0$, regardless of the initial topology $\nu$.  Other discretized flow equations~\cite{Braam:1988qk, GarciaPerez:1989gt, deForcrand:1997esx} may behave more like the continuum case \eq{flow}. Therefore the two cases of greatest interest to us will be $\nu_\star=\nu$ (the flow equation preserves topology), and $\nu_\star=0$ (the flow equation completely destroys topology).\footnote{For a discussion of topology on the lattice and how flow equations derived from various actions will affect topological charge, see Ref.~\cite{GarciaPerez:1993lic}.}

\section{Review of the derivation of vector overlap operator}
\label{OverlapFormalism}

The two salient features of the conventional domain wall construction in the $L\to \infty$ limit are (i) the correct realization of anomalies via the Chern-Simons term due to non-decoupling of the bulk fermions, and (ii)  the protection of fermion mass terms from additive radiative corrections due to the localization of the positive and negative chiral modes far from each other in the extra dimension.  However neither feature can be dealt with simply or rigorously in the five-dimensional formulation.  A truly four-dimensional lattice description of this theory was derived in a series of brilliant papers by Narayanan and Neuberger who realized that the domain wall partition function in the $L\to\infty$ limit can be described in terms of the overlap of vacua of two different four-dimensional Hamiltonians~\cite{Narayanan:1994gw, Narayanan:1993sk}.
With an expression for the fermion determinant in hand as a guide, they were then able to construct the fermion kinetic operator (known as the overlap operator) and show that it solves the Ginsparg-Wilson (GW) equation.  L\"uscher then showed that simply by being a solution of the GW equation the overlap operator correctly reproduces the index theorem relating fermion zeromodes to gauge topology, accounting for the physics of the Chern-Simons operator in the five-dimensional formulation~\cite{Luscher:1998pqa}. He also showed that the overlap operator respects an exact $U(1)$ symmetry, even at finite lattice spacing, ensuring that fermion masses could only be multiplicatively renormalized. This was the final piece of the puzzle to explain how domain wall fermions in the limit $L \to \infty$ correctly regulate massless Dirac fermions.   Here we review the construction of the conventional overlap operator for the vector-like theory, and then apply the same analysis to the chiral theory.

Our starting point is the Shamir form of the five-dimensional  lattice theory on a slab  to better make connection with much of the literature on overlap fermions, in particular Refs.~\cite{Vranas:1997da,Neuberger:1997bg,Kikukawa:1999sy}. 
The  domain wall theory for vector-like gauge theories is pictured in Fig.~\ref{fig:vcoverlap}, corresponding to  the lattice action
\beq
S = \sum_{x,s} \bar\psi\left[-P_- \nabla_5 +P_+ \nabla_5^*  +\gamma_5 H\right]\psi
\eeq
where 
\beq
 \gamma_5 H &= &\half\left[ \gamma_\mu(\nabla_\mu + \nabla_\mu^*) - \nabla_\mu\nabla_\mu^*\right]-m\cr&&\cr &\equiv& D_w-m
\eqn{ham}\eeq
with $m$ is the fermion mass, $\mu = 1,\ldots,4$, and
\beq
P_\pm = \frac{1\pm\gamma_5}{2}\ .
\eeq
The coordinate runs over $s=0,\ldots,L$, the $\gamma_\mu$ are hermitian, and we have set the lattice spacing and fermion mass to $a_5=a=1$. The forward and backward lattice covariant derivatives  are $\nabla$ and $\nabla^*$  respectively, defined as
\beq
\nabla_\alpha \psi_\bfn &=& U_\alpha(\bfn) \psi_{ \bfn+\hat \alpha}-\psi_\bfn\ ,\cr &&\cr
\nabla^*_\alpha \psi_\bfn &=& \psi_\bfn - U^\dagger _\alpha(\bfn-\hat \alpha) \psi_{ \bfn-\hat \alpha}\ ,
\eeq
 where $U_\alpha$ are the $s$-independent gauge links in the $\hat a$ direction with $U_5=1$. The fermions  satisfy the boundary conditions
\beq
P_+\psi(x,0) = P_-\psi(x,L)=0\ .
\eeq
 Requiring that the negative chirality fermion be localized near $s=0$ bounds the fermion mass to lie in the range $1\geq m>0$. We choose $m =1$ so that the negative chirality fermion is confined to sit at the $s= 0$ slice; similarly, the positive chirality fermion is confined to $s=L$.  
 
 The spectrum of this theory includes in the large $L$ limit a massless negative chirality fermion bound to the surface at $s=0$, and a massless positive chirality fermion bound to the surface at $s=L$, where the mass vanishes exponentially fast in $L$. In addition to the fermions, Pauli-Villars fields with anti-periodic boundary conditions at $s=0,L$ but identical action are also needed \cite{Vranas:1997da}. Considering the discrete $s$ coordinate as a flavor index, the $L\to\infty$ limit corresponds to $N_f\to\infty$, and the role of the Pauli-Villars fields, first introduced in Ref.~\cite{ Narayanan:1993sk}, is to cancel contributions from the bulk fermions that would lead to divergences in this limit.

The effective kinetic operator for the surface modes was computed in Ref.~\cite{Kikukawa:1999sy} using techniques developed in Refs.~\cite{Narayanan:1994gw,Neuberger:1997bg} for computing the fermion determinant, with the result that\footnote{We use a canonically normalized $\hat\CD_\chi$, differing from  Ref.~\cite{Kikukawa:1999sy} by a factor of 2.}
\beq
\CD_V =2\lim_{L\to\infty} \frac{\CD_f}{\CD_{PV}} \ ,
\eqn{rat}
\eeq
where $\CD_f$, $\CD_{PV}$ are the fermion and Pauli-Villars contributions respectively, computed to be (up to common factors)
\beq
\CD_f = \frac{1+\gamma_5 \tanh\frac{L}{2} H}{1-\gamma_5 \tanh\frac{L}{2} H}\ ,\qquad \CD_{PV} = \CD_f + 1\ .
\eqn{fpv}
\eeq
The large $L$ limit can be taken by defining a matrix $\CE^{(L)}_V$ in terms of the transfer matrix $T=\exp{(- H)}$ for propagation in the fifth dimension
\beq
\CE^{(L)}_V =\frac{1 - T^{L}}{1+ T^{L}} = \tanh\frac{L}{2} H \ .
\eqn{CEdefV}
\eeq
 The large $L$ limit of $\CE^{(L)}_V$ is 
\beq
\lim_{L \to \infty}\CE^{(L)}_V = \epsilon(H)
\eeq
where  $\epsilon$ is the sign function which can be represented as
\beq
\epsilon(h) = \frac{h}{\sqrt{h^2}}
\eeq
for a Hermitian matrix $h$. Combining the above equations  leads to the result
\beq
\CD_V &=&  \lim_{L \to \infty}\left( 1 + \gamma_5 \CE^{(L)}_V\right) 
=   1 + \gamma_5 \epsilon\ ,
\eqn{OverlapV}
\eeq
where we adopt the abbreviation $\epsilon\equiv \epsilon( H)$.  While the Hamiltonian defined via the transfer matrix is not the same as the Hamiltonian in \eq{ham}, the latter can be used for defining $\epsilon$ in the $L\to \infty$ limit.
This is the standard overlap operator that solves the GW equation \cite{Neuberger:1997fp,Neuberger:1998wv};  it solves the GW equation by virtue of the fact that $\epsilon^2=1$. The important role of $\epsilon$ was already recognized in Ref.~\cite{ Narayanan:1993sk}. 

\section{Derivation and properties of a chiral overlap operator}
\label{chiprops}

It is straightforward to generalize the derivation of $D_V$ given in the previous section to the problem of interest here, namely the marriage of domain wall fermions with gradient flow for the gauge field, pictured in the middle panel of Fig~\ref{fig:vcoverlap}, where the negative chirality component interacts with an arbitrary gauge field $A$, while the positive chirality component sees a gauge field $A_\star$.   For general flow we can simply replace $T^L$ in \eq{CEdefV} by
\beq
T^L \to \prod_{s=L}^1 T(s) = \prod_{s=L}^1 e^{-H(s)}
\eqn{grad}\eeq
where the $s$-dependence of the transfer matrix is solely due to the $s$-dependence of the flowing gauge field.  The formal expression for the chiral overlap operator is therefore
\beq
\CD_\chi = 1 + \gamma_5 \CE_\chi \ ,
\eqn{res1}\eeq
where
\beq
 \CE_\chi& =& \lim_{L\to\infty} \CE^{(L)}_\chi\ ,\cr 
 \CE^{(L)}_\chi &=&\frac{1 -  \prod_s  T(s)}{1+  \prod_s  T(s)} = \tanh\left[\frac{L}{2} \log\left(\prod_{s } T(s)\right)^{\frac{1}{L}}\right] \ .
 \eqn{res2}\eeq
and the logarithm is well-defined as $T(s)$ is positive. If one regards the logarithm in the above expression as a sort of averaged Hamiltonian with finite eigenvalues in the large $L$ limit then $ \CE_\chi^2 = 1$, which ensures that $D_\chi$ obeys the Ginsparg-Wilson equation. This is analagous to the vector case where $\CE_V = \epsilon(H)$.

To allow for explicit analysis we consider the special case shown in the right panel of Fig~\ref{fig:vcoverlap}. The flow time $\tau(s)$ in \eq{flow}  is a function of $s$ such that the gauge field $A$ remains roughly constant over a region of $O(L)$ near the $s=0$ boundary, then makes a quick transition to the fixed point gauge field $A_\star$, which remains roughly constant over a region of size $O(L)$ near the $s=L$ surface. There are potential problems with this simplification which we will address below. The simplification  allows us to construct the corresponding chiral overlap operator $\hat\CD_\chi$ by simply replacing
\beq
T^L \to T_\star^{L/2}T^{L/2}
\eqn{sudden}\eeq 
in \eq{CEdefV} before taking the $L\to\infty$ limit. The transfer matrices $T$ and $T_\star$  corresponding to Hamiltonians with gauge fields $A$ and $A_\star$ respectively. Therefore we have
\beq
 \hat\CD_\chi =  \left( 1 + \gamma_5 \hat\CE_\chi\right) \ .
\eqn{result}\eeq
where the hats on $\hat \CD_\chi$ and  $ \hat\CE_\chi$ signifies the assumption of an abrupt transition from $A$ to $A_\star$, distinguishing them from the more general formulas of \Eqs{res1}{res2}.  We see that $\hat \CE_\chi$ is defined as the limit
\beq
\hat\CE_\chi  &=&  \lim_{L\to\infty}  \hat\CE_\chi^{(L)}=\lim_{L\to\infty} \left(\frac{1 - T_\star^{L/2}T^{L/2}}{1+ T_\star^{L/2}T^{L/2}}\right)
\ .
\eqn{CEdefC}
\eeq
The matrix $ T_\star^{L/2}T^{L/2}$ in the above expression has positive definite real eigenvalues, being the product of two positive semi-definite hermitian matrices.  Thus the eigenvalues of the matrix $\hat\CE_\chi^{(L)}$ are bounded to lie in the interval $(-1,1)$.  Since the eigenvalues of both $ T_\star^{L/2}$ and $ T^{L/2}$ become either zero or infinite in the infinite $L$ limit, the eigenvalues of $ T_\star^{L/2}T^{L/2}$ will also become either zero or infinite in this limit except at possible exceptional gauge field configurations for which the spectra of $H$ and $H_\star$ are related.  This implies that the eigenvalues of $\hat\CE_\chi$ will equal $\pm1$. As an example of how this conclusion could fail, suppose there exists a vector $\psi$ satisfying $H\psi = -H_\star\psi = E\psi$; in this case  $\hat\CE_\chi$ has a zero eigenvalue. This will not happen for generic gauge fields, nor will it occur in perturbation theory.

A more useful form for $\hat\CE_\chi$ can be found by making use of the limits
\beq
 \lim_{L\to\infty} \frac{1 - T^{L}}{1+ T^{L}}= \epsilon\ ,\qquad 
 \lim_{L\to\infty}  \frac{1 - T_\star^{L}}{1+ T_\star^{L}} = \epsilon_\star 
\eeq
where 
\beq
\epsilon \equiv \epsilon(H[A])\ ,\qquad \epsilon_\star \equiv  \epsilon(H[A_\star])\ ,
\eqn{epsdef}\eeq
and   $H$ is the Wilson   Hamiltonian in \eq{ham}.   By manipulating the matrices carefully at finite $L$ before taking the $L\to\infty$ limit we arrive a central  result of this paper, 
\beq
\hat\CE_\chi = \left[1-\left(1-\epsilon_\star \right)\frac{1}{1+\epsilon\, \epsilon_\star}\left(1-\epsilon\right)\right]\ ,
\eqn{ChiralOverOp}
\eeq
with chiral overlap operator $ \hat\CD_\chi = 1+ \gamma_5 \hat\CE_\chi $.

The above expression requires some care in its interpretation since the denominator $(1+\epsilon\, \epsilon_\star)$ can have zero eigenvalues.  In fact, as we show in the appendix, the denominator has zero eigenvalues whenever gradient flow destroys topology. However, the numerator also vanishes in this case and so \eq{ChiralOverOp} is consistent with the bound we have placed on the eigenvalues of $\hat\CE_\chi$.  We return to this case in the next section but first we exhibit the key properties of $\hat\CD_\chi$ for the case $\nu=\nu_\star$, where $\nu$ and $\nu_\star$ are integers defined as
\beq
\nu\equiv \half \Tr\epsilon \ , \qquad \nu_\star\equiv \half \Tr\epsilon_\star \ ,
\eqn{nudef1}
\eeq
and both $\epsilon$ and $\epsilon_\star$ have even dimension. In the continuum limit, $\nu$ and $\nu_\star$ can be identified as the winding number in the gauge fields $A$, $A_\star$ respectively.  For the case  $\nu=\nu_\star$ that we are considering here,  $1+ \epsilon \epsilon_\star$ is generically invertible and $\hat\CE_\chi$ well-defined.

\subsection{The continuum limit of $\hat\CD_\chi$}

In order for $\hat\CD_\chi$ to be the Euclidean fermion operator for Weyl fermions, it must have the expected continuum limit, i.e. the negative and positive chirality Weyl fermions must decouple. The continuum limit should be as given in \eq{AGchi}.  The operator $\epsilon(H)$ has the continuum expansion
\beq
\epsilon &=& \gamma_5\left(-1+ \slashed{D}(A)+O(a)\right)\ ,\cr
\epsilon_\star &=& \gamma_5\left(-1+  \slashed{D}(A_\star)  +O(a)\right)\ ,
\eeq
where we have set $m = 1$. For the vector theory this expansion leads to the continuum limit 
\beq
\CD_V = \slashed{D}+ O(a)\ ;
\eeq
the proper normalization occurs because with $m=1$, the zeromodes are completely confined to the four-dimensional surfaces at $s=0$ and $s=L$ to leading order in perturbation theory.    Performing the same expansion for $\hat\CD_\chi$ in \eq{ChiralOverOp} we find 
 \beq
 \hat\CD_\chi = \begin{pmatrix}0&  \sigma_\mu D_\mu(A) \\ \bar\sigma_\mu D_\mu(A_\star) &0\end{pmatrix} + O(a)\ .
\eqn{cl}\eeq
This result is then identical  to  \eq{AGchi}, and confirms that in the continuum limit, the negative chirality zeromode sees a field $A$ while its positive chirality partner sees the flowed field $A_\star$.
This is  a desirable result, but in itself is insufficient to show that the mirror fermions decouple in the continuum, as it is a tree level calculation.  One would like to see the determinant factor into fermion and mirror contributions in the continuum limit, but being a loop calculation, the subleading terms dropped in \eq{cl} can spoil the desired factorization.

\subsection{ $\hat\CD_\chi$ satisfies the Ginsparg-Wilson equation}
\label{gweq}

Any operator of the form
\beq
\CD = 1+ \gamma_5 \CE
\eeq
satisfies the Ginsparg-Wilson equation if $\CE^2=1$. For the vector overlap operator, 
\beq
\CE_V =\epsilon
\eeq
and  so $\CE_V^2=1$. For the chiral overlap operator, we have already argued that except for possible exceptional gauge fields, the eigenvalues of $ \CE_\chi$ equal $\pm1$, assuring that $ \CD_\chi$ also satisfies the GW equation.  For  an explicit calculation using the expression in \eq{ChiralOverOp} we can write
\beq
\hat\CE_\chi&=& \left[1-\left(1-\epsilon_\star \right)\frac{1}{1+\epsilon\, \epsilon_\star}\left(1-\epsilon\right)\right]\cr &&\cr
&=&
 \left[\epsilon+\left(1+\epsilon \right)\frac{1}{\epsilon+ \epsilon_\star}\left(1-\epsilon\right)\right] \ ,
\eqn{Esimp}\eeq
and in the latter form one sees that the second term does not contribute to $\hat\CE_\chi^2$ and so it  follows immediately that $\hat\CE_\chi^2=1$. Therefore  $\hat\CD_\chi$ satisfies the GW equation.  The chiral solution is also a generalization of the vector solution, since for the special case $A_\star\to A$ one has $\epsilon_\star\to \epsilon$ and   $\hat\CD_\chi\to D_V$, as is evident  from the five-dimensional domain wall construction when gradient flow is turned off.
 
 %%%%%%%%%%
\begin{figure}[t]
\includegraphics[width=5 cm]{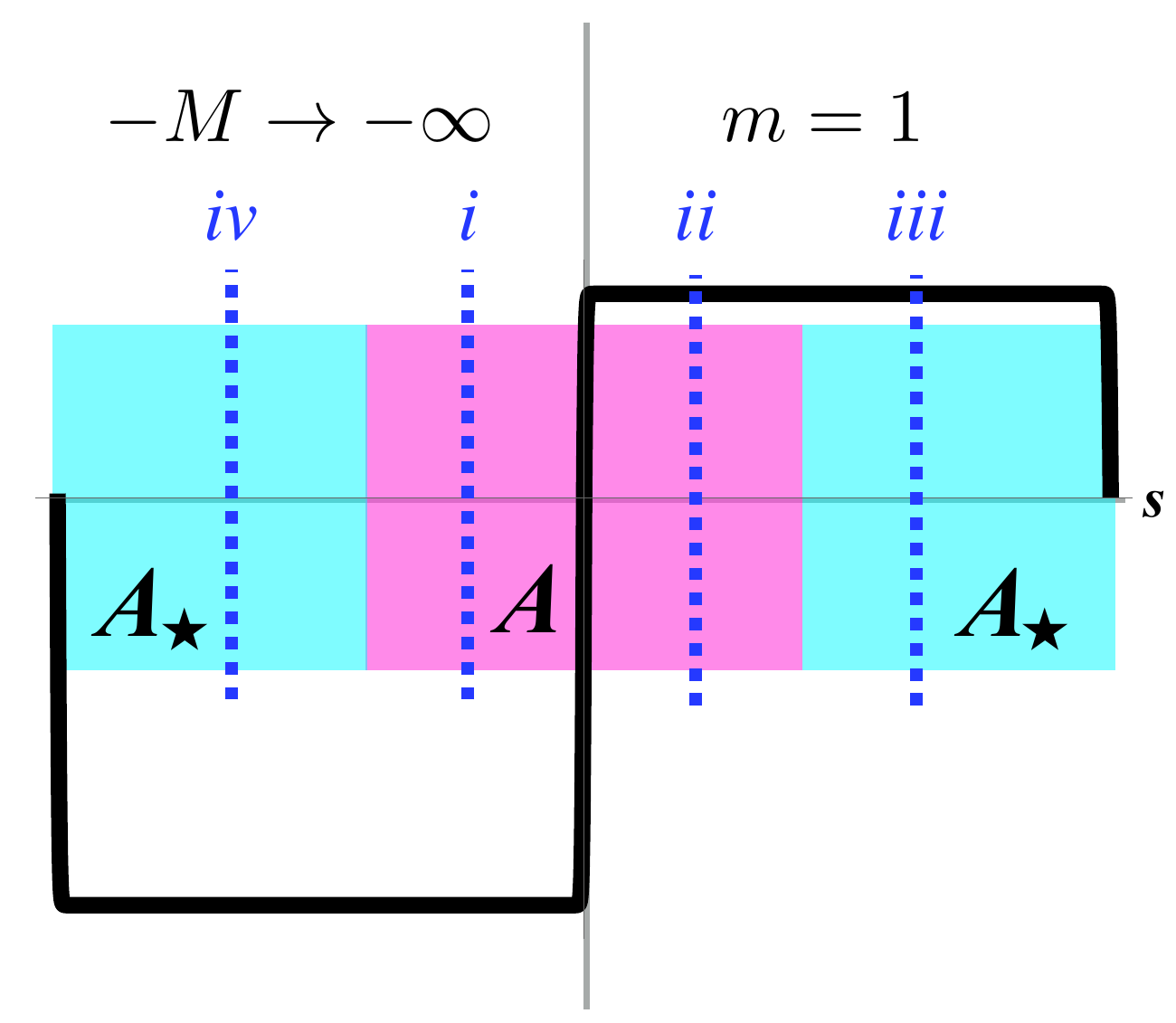}
\caption{  \it  The four regions in the domain wall construction of Ref.~\cite{Grabowska:2015qpk}.}
\label{fig:vacover}
\end{figure}
%%%%%%%%%%%%

\subsection{ $\hat\CD_\chi$ does not suffer from phase ambiguity}

To see that there is no phase ambiguity in the definition of the determinant of $\hat\CD_\chi$, it is convenient to consider the five-dimensional domain wall construction of Ref.~\cite{Grabowska:2015qpk} shown in   Fig.~\ref{fig:vacover}, with  a Dirac fermion living in the space $s\in [-L,L]$ with $s=\pm L$ identified.  The fermion mass changes sign across the defects at $s=0$ and $s=\pm L$, while the gauge field equals $A$ near the $s=0$ defect and $A_\star$ near the $s=L$ defect. Two Pauli-Villars fields can be introduce to eliminate bulk effects, one with mass $M$ on $s\in [-L,0]$  and the other with mass $-m$ on $s\in [0,L]$, both  with   anti-periodic boundary conditions in $s$. Thus there are four distinct regions  in $s$ corresponding to the two values for the fermion mass and the two values of the gauge field, described by the Hamiltonians 
\beq
H_1 &=&\gamma_5\left( D_w(A)+ M\right) \ ,\cr 
H_2&=&\gamma_5 \left( D_w(A)- m\right) \ ,\cr 
H_3 &=&\gamma_5 \left( D_w(A_\star)- m \right)\ ,\cr 
H_4 &=&\gamma_5 \left( D_w(A_\star)+M\right) \ ,
\eqn{chiham}\eeq
where $D_w$ is the Wilson operator given in \eq{ham}, and we will take $M\to\infty$ and set $m=1$ in lattice units.

In the original Narayanan-Neuberger derivation, there is no gradient flow and so $A_\star = A$. Inserting a complete basis of states at two
of the four s-slices (i and ii in  Fig.~\ref{fig:vacover}) and taking the large $L$ limit, with the distances between all the relevant surfaces becoming infinite, all excited energy states are projected out and the partition function can be expressed in terms of the many-body ground states of the Hamiltonians $H_n$, denoted $\ket{\Omega_n}$, as 
\beq
\det\CD_V= \frac{\braket{\Omega_1|\Omega_2} \braket{\Omega_2|\Omega_1} }{\braket{\Omega_1|\Omega_1} \braket{\Omega_2|\Omega_2}  }\ ,
\eqn{vacoV}
\eeq
where the numerator arises from the fermions while  the denominator is due to the Pauli-Villars fields.   The $\ket{\Omega_n}$ ground states in each case correspond to a filled Dirac sea and their overlap can be expressed as determinants of the overlap of negative energy single-particle wavefunctions.  

Extending the  Narayanan-Neuberger procedure to the more complicated system we are interested in with $A_\star \neq A$, we insert a complete set of states at all four $s$-slices  (i-iv) shown in Fig.~\ref{fig:vacover} and take the $L\to\infty$ limit. Again, the partition function can be expressed in terms of the $\ket{\Omega_n}$ states as
\beq
\det\hat\CD_\chi= \frac{\braket{\Omega_1|\Omega_2} \braket{\Omega_2|\Omega_3} \braket{\Omega_3|\Omega_4} \braket{\Omega_4|\Omega_1} }{\left\vert\braket{\Omega_2|\Omega_3}\right\vert^2 \left\vert \braket{\Omega_1|\Omega_4} \right\vert^2 }\ .
\eqn{vaco}\eeq
While \eq{vaco} is not particularly useful for deriving  $\hat\CD_\chi$, the cyclic arrangement of the four $\ket{\Omega_n}\bra{\Omega_n}$ factors in the numerator -- which follows from the compact nature of the fifth dimension in the domain wall formulation --  makes this expression for the chiral partition function both manifestly complex and  independent of any phase convention for the ground states  $\ket{\Omega_n}$.  This is to be contrasted with earlier attempts to define the chiral determinant on a non-compact fifth dimension  \cite{Kaplan:1992bt,Narayanan:1993sk} which suffered the same phase ambiguity encountered in  defining the chiral fermion determinant in the four-dimensional continuum, namely that the right and left Hilbert spaces are subject to independent phase conventions.

\section{The Index of the chiral overlap operator}
\label{Index}
The $U(1)_A$ anomaly for the vector overlap operator can be expressed by the index theorem
\beq
-\half\Tr\gamma_5 \CD_V  =-\nu =   (N_+-N_-)\ ,
\eqn{anomalyV}\eeq
where $N_\pm$ are the number of exact zeromodes of $\CD_V$ and $\nu$ is defined in \eq{nudef1}.  Only the fact that $\CD_V$ satisfies the GW equation is required to prove this relation \cite{Luscher:1998pqa}. In the continuum limit and for smooth gauge fields, the integer $\nu$ as calculated from $\epsilon$ can be equated to the usual topological winding number of the gauge fields, proportional to $\int \Tr F \widetilde F$. In this case \eq{anomalyV} coincides with the continuum index theorem for a Dirac fermion~\cite{Fujikawa:1998if, Adams:1998eg, Suzuki:1998yz}. 

To understand what to expect for the index equation for $\CD_\chi$ analogous to \eq{anomalyV} we first consider the continuum operator \eq{AGchi}.  From the properties of the Dirac operator in a background gauge field $A$ with winding number $\nu$, we know that there are $\half(|\nu|+\nu)$ normalizable 2-component solutions to the differential equation
\beq
D_\mu(A) \sigma_\mu \phi = 0\ ,
\eeq
and  $\half(|\nu|-\nu)$ normalizeable solutions to 
\beq
(D_\mu(A))^\dagger  \sigma_\mu \phi =0\ .
\eeq
Here we are ignoring ``accidental zeromode" solutions for special gauge field which are not mandated by topology.    Analogous equations hold substituting $A\to A_\star$ and $\nu\to\nu_\star$.  Since $\bar\sigma_\mu = \sigma_\mu^\dagger$ if follows that $\CD_\chi$ in \eq{AGchi} has $n_\pm$ right zeromodes with chirality $\pm1$ and $\bar n_\pm$ left zeromodes with chirality $\pm1$ where
\begin{equation}
\begin{aligned}
  n_+ &= \frac{|\nu_\star|-\nu_\star}{2}\ ,\quad
&n_- &= \frac{|\nu|+\nu}{2} \ ,\cr 
  \bar n_+ &= \frac{|\nu|-\nu}{2}\ ,  
& \bar n_- &= \frac{|\nu_\star|+\nu_\star}{2}\ .
\end{aligned}
 \end{equation}
giving rise to nonzero matrix elements of the generic form
\beq
\vev{{\bar\psi}^{\bar n_-}_R {\bar\psi}^{\bar n_+}_L \psi^{n_+}_R\psi^{n_-}_L}
\eeq
which violates the chiral charge by
\beq
\Delta Q_5  &=&  -\left( \underbrace{(n_-- \bar n_+ )}_{\nu}  + \underbrace{(\bar n_- - n_+)}_{\nu_\star}\right)\nonumber \\
&=&  -(\nu + \nu_\star)\ .
\eeq
Thus we expect the analogue to \eq{anomalyV} for the chiral operator should be
\beq
- \Tr\gamma_5 \CD_\chi  =-(\nu+\nu_\star) =   2(N_+-N_-)\ ,
\eqn{contChiralindex}\eeq
where
\beq
n_- = \bar n_-  = N_-\ ,\quad n_+= \bar n_+ = N_+\ .
\eeq

%To understand what may be expected for a continuum chiral theory with the fermion operator of \eq{AGchi}, it is useful to first solve the anomaly equation \eq{anomalyV} for $n_\pm$ in the vector case. Utilizing the constraint that $n_\pm$ are non-negative integers,  its solution is
%\beq
%n_- = \half (|\nu|+\nu) +p\,, \quad n_+ =  \half (|\nu|-\nu) +  p \, ,
%\eeq
%where $p$ is an arbitrary non-negative  integer counting ``accidental" zeromodes.  Generalizing to the case where the negative chirality fermion only sees the gauge field $A$ while the positive chirality fermion only sees the gauge field $A_\star$, the solutions for $n_\pm$ should be
%\beq
%n_- = \half (\nu+|\nu|) +p\, ,\quad n_+ =  \half (\nu_\star-|\nu_\star|) +  p_\star \, ,
%\eeq
%where $p$ and $p_\star$ are now independent non-negative integers. The difference gives the generalized version of the index theorem
%\beq
%\half\left[(\nu+\nu_\star) + |\nu|-|\nu_\star| + 2q\right] =(n_- - n_+)\ ,
%\eqn{contChiralindex}
%\eeq
%where $q=(p-p_\star)$ is an arbitrary integer. Unlike in the vector case, there is no strict relation between the winding numbers $\nu$, $\nu_\star$ and the number of zero modes, $(n_--n_+)$.  The numbers $p$ and $p_\star$  count independent ``accidental" zeromodes in the $A$ and $A_\star$ gauge fields which are not mandated by topology.

We now compute the lattice version of  \eq{contChiralindex} directly from our expression for  $\hat \CD_\chi$, understood as the limit in \eq{CEdefC}. Since $\hat \CD_\chi$ obeys the GW equation, it too obeys the chiral anomaly equation
\beq
-\Tr\gamma_5 \hat\CD_\chi = -\Tr\hat\CE_\chi = 2 (N_+-N_-)\ ,
\eqn{anomalyX}
\eeq
with 
\beq
 \hat\CE_\chi =\left[1-\left(1-\epsilon_\star \right)\frac{1}{1+\epsilon\, \epsilon_\star}\left(1-\epsilon\right)\right]\ .
\eeq
What remains to do, then, is to show that $\Tr\hat\CE_\chi = (\nu+\nu_\star)$.  This is easy to do for the case of topology preserving flow, where $\nu=\nu_\star$.  As discussed in the appendix, for this case $(1+\epsilon\, \epsilon_\star)$ is invertible and the trace can be trivially computed using the form in \eq{Esimp} with the desired result that
%\beq
%\Tr\hat\CE_\chi = \Tr \epsilon = 2\nu = (\nu+\nu_\star)\qquad \text{($\nu=\nu_\star$)}\ .
%\eeq
\beq
\Tr\hat\CE_\chi  = 2\nu \qquad \text{for}\, \nu=\nu_\star\ .
\eeq

For unconstrained $\nu$ and $\nu_\star$, relevant for when the lattice gradient flow equation causes topology to vanish, the analysis is more complicated since $(1+\epsilon\, \epsilon_\star)$ is no longer invertible and $\hat\CE_\chi$ must be defined as the limit in \eq{CEdefC}.  
 In \eq{TrEX}  this trace is computed with the result
 \beq
\Tr\hat\CE_\chi = (\nu+\nu_\star)  + \sum_{i=1}^{|\nu-\nu_\star|} \xi_i \ ,\qquad \xi_i = \pm1\ .
\eqn{TrEX2}
\eeq
In the above expression the first term is the contribution from eigenstates of $U= \epsilon\epsilon_\star$ with eigenvalue $+1$, and the second term is the contribution from eigenstates of $U$ with eigenvalue $-1$. This second term vanishes when the flow preserves topology and $\nu=\nu_\star$ .   The $\xi_i$ parameters cannot be determined solely in terms of  $\nu$ and $\nu_\star$ as they depend on the interplay between the two gauge fields; they must take on values of $\pm1$. While $U$ also has eigenstates with generally complex eigenvalues, they are shown not to contribute to the trace.  We see that for cases where $|\nu-\nu_\star|$ is an even integer it might be possible for the $\xi_i$ sum to vanish, recovering the correct continuum anomaly \eq{contChiralindex}; however for odd $|\nu-\nu_\star|$ this is never possible.  

The conclusion of this section is that when we assume a rapid transition from $A$ to $A_\star$ in the bulk, obtaining the correct index theorem requires  topology preserving gradient flow on the lattice. 

 \section{Problems with sudden flow}
 
 In order to be able to treat the derivation of the chiral overlap operator analytically we have chosen to consider the sudden flow scenario pictured in the third panel of Fig.~\ref{fig:vcoverlap}, instead of the gradual flow shown in the middle panel.  This allowed us to replace the general form for the product of transfer matrices in \eq{grad} with the more tractable form in \eq{sudden}.  From the five-dimensional picture, however, the decoupling of fermions on one surface from their mirror partners on the other relied on the fermions having purely local interactions with a mass gap in the bulk.  In effect the sudden transition from $A$ to $A_\star$ in the middle of the sample implies that a fermion in the bulk can couple simultaneously to the gauge fields at both boundaries.  On integrating out the bulk fermion one can then in principle generate gauge invariant operators which are functions of the difference $(A-A_\star)$, an object that transforms homogeneously under gauge transformations.

 An example of such an unwanted term for the theory considered here in the sudden gauge field flow  scenario was found in Ref.~\cite{Makino:2016auf}, which computed the anomaly for smooth gauge fields.  The authors of that paper found that the divergence of the fermion current included a Lorentz-violating operator of the form
 \beq
\sum_\mu \partial_\mu \Tr C^3_\mu\ ,
 \eeq
 where $C  = (A_\star-A) $.  This result implies that $\nu$ and $\nu_\star$ are not given by the standard gauge field winding number in the continuum limit. This operator is seen to vanish under the same condition that ensures cancellation of gauge anomalies, namely that the symmetrized trace of three gauge generators vanishes, $\Tr T_a\{T_b,T_c\}=0$.  However, chiral gauge theories will contain other $U(1)$ charges for which such a term will not vanish.  A simple example is an $SU(5)$ chiral gauge theory with Weyl fermions transforming as $\bar 5 \oplus 10$.  Such a theory has no $SU(5)$ gauge anomaly; it also has an anomaly-free global $U(1)$ symmetry where the $\bar 5$ carries charge $Q=3$ and the $10$ carries charge  $Q=-1$.  The results of Ref.~\cite{Makino:2016auf} imply that computation of the divergence of this current on the lattice  using the sudden flow result  \eq{ChiralOverOp} will result in a term of the form 
 \beq
\sum_\mu \partial_\mu \Tr QC^3_\mu\ 
 \eeq
which will not vanish.  This result, which persists in the continuum limit, is incompatible with the fermion determinant successfully factorizing into fermion and mirror contributions.  From the five-dimensional picture we do not expect such terms to exist when gradual gauge flow is used, as in \eq{grad}, but that remains to be shown.\footnote{Our discussion in this section has benefitted greatly from comments by M. L\"uscher and E. Poppitz.}

 \section{Discussion}
  \label{Discussion}
 
 We have considered here the lattice formulation of the proposal in Ref.~\cite{Grabowska:2015qpk} for combining five-dimensional domain wall fermions with gradient flow for the gauge fields. Our proposal  \Eqs{res1}{res2} for a chiral fermion operator follows a venerable list of such proposals in the literature that have succumbed for various reasons. Therefore it is important to explore this new one further, both analytically and numerically, to test its viability.  In particular it is important to understand better the issue of factorization of the fermion determinant in the continuum limit. One possibly interesting avenue for numerical exploration is the role of topologically induced interactions between matter and fluff,  as discussed briefly in Ref.~\cite{Grabowska:2015qpk}, and whether they persist in the large volume limit.  These questions can most likely be explored  by constructing $D_\chi$ for two fermions in conjugate representations of the gauge group, so that the ``chiral" gauge theory being considered is actually vector-like. Choosing a theory such as QCD, which has been well-studied on the lattice, allows for comparison with conventionally obtained results and is likely to not be afflicted by a sign problem, unlike less trivial chiral gauge theories.   
 
In order to treat $D_\chi$ explicitly we considered the particular choice of abrupt flow from $A$ to $A_\star$, pictured in the third panel of  Fig.~\ref{fig:vcoverlap}, for which the expression $\hat D_\chi$ was derived in \eq{result}.  This allowed us to illustrate explicitly that $\hat \CD_\chi$ obeyed the GW equation and, for topology preserving flow, gave the relation between the its index and the topological index $\nu$.  However, in relating $\nu$ to the winding number of smooth gauge fields,  the results of Ref.~\cite{Makino:2016auf} exhibit a pathological dependence on the gauge fields which indicates that fermions and their mirrors do not successfully decouple from each other.  We therefore believe that the gradual flow form given by \eq{grad} is the proper formulation of our idea, although we do not have a simple analytical form for the fermion operator in this case. Since our conclusion that the flow has to be topology preserving is only based on the abrupt flow scenario, we remain agnostic about whether or not a topology changing flow equation on the lattice must be ruled out entirely.
  
  It is also necessary  to better understand better  the role of gauge anomalies, a topic not addressed in this paper.  A feature of our proposal is that it is manifestly gauge invariant, while a chiral gauge theory with an anomalous fermion representation in the continuum is not. In the five-dimensional domain wall construction of the theory,  models which do not have anomaly cancellation at each defect are characterized by a nonzero coefficient for a bulk Chern-Simons operator, which is nonlocal due to the gradient flow of the gauge field~\cite{Grabowska:2015qpk}.   We have not explored here the consequences of anomalous fermion representations for our four-dimensional chiral overlap operator $\hat\CD_\chi$.  The subject of global anomalies has also been recently explored for domain wall fermions with gradient flow in Ref.~\cite{Fukaya:2016ofi}, but how that physics manifests itself in the chiral overlap operator remains to be explored. 
  
Finally, it would be gratifying to be able to establish a connection  between the chiral overlap operator derived here, and  L\"uscher's   lattice construction of Abelian chiral gauge theories and consistency conditions on nonabelian ones  presented in Refs.~\cite{Luscher:1999un,Luscher:1998du}. 
 
 %\vfill
 
\begin{acknowledgments}
We received many useful criticisms of an earlier version of this work which we have now addressed. M. L\"uscher pointed out problems with the assumption of abrupt flow from $A$ to $A_\star$ that we could treat analytically, but which will interfere with the decoupling of the mirror fermions; M. Golterman and Y. Shamir found an error in our original treatment of the index of the continuum operator in \eq{cl}; E. Poppitz showed us that the result of Ref.~\cite{Makino:2016auf} gave evidence for non-decoupling of the mirror fermions, as expected from M. L\"uscher's comments, when considering the divergence of anomaly-free $U(1)$ currents in theories with gauge anomaly cancellation. We gratefully also acknowledge communications related to this work with H. Fukaya, M. Garcia Perez, A. Hasenfratz, H. Onogi,  and H. Suzuki.  DMG would also like to acknowledge the hospitality of the University of Colorado Physics Department where part of this work was completed.  This  work was supported by the NSF Graduate Research Fellowship under Grant No. DGE-1256082,  the DOE Grant No. DE-FG02-00ER41132 and the NSF Grant No. 82759-13067-44-PHHXM.  
\end{acknowledgments}

\begin{appendix}
\label{Appendix}

\section{Properties of $ \epsilon\epsilon_\star$}
Here we discuss some properties of  the matrix $U \equiv  \epsilon\epsilon_\star$,  where $\epsilon$ and $\epsilon_\star$ are the sign functions of two independent $2N\times 2N$ hermitian matrices $H$ and $H_\star$, 
\beq
\epsilon = \frac{H}{\sqrt{H^2}}\ ,\qquad \epsilon_\star =  \frac{H_\star}{\sqrt{H_\star^2}}\ ,
\eeq
with traces given by
\beq
\Tr\epsilon = 2\nu\ ,\qquad \Tr\epsilon_\star = 2\nu_\star\ .
\eqn{nudef}
\eeq
The properties of 
\beq
U\equiv  \epsilon\epsilon_\star
\eqn{Udef}
\eeq
 are important for understanding our expression for $D_\chi$ as it involves the inverse of $\Delta^+ = (1+U)$; these properties are examined in this section.

\begin{lemma}
Eigenvalues of $U$ take the values of $\pm1$, or come in complex conjugate pairs, where $\epsilon$ is the conjugation matrix.
 \end{lemma}
 
 \smallskip
 \noindent
 As matrix $U$ is unitary its eigenvalues are phases;  left and right eigenvectors are conjugates of each other and we can represent them using Dirac's bra-ket notation:
\beq
U\ket{m} = \eta_m\ket{m}\ ,\quad \bra{m} U = \eta_m\bra{m} \ 
\eeq
with $|\eta_m|^2=$ and $\braket{m|n} = \delta_{mn}$.  The state $\ket{m}$ must also satisfy
\beq
U^\dagger \ket{m} = \eta_m^*\ket{m}\ .
\eeq
 Since $U=\epsilon\epsilon_\star$ and $U^\dagger=\epsilon_\star\epsilon$,
 \beq
 U \epsilon \ket{m} = \epsilon U^\dagger\ket{m} = \eta_m^* \epsilon \ket{m} \ .
\eqn{conj}\eeq
Therefore the complex eigenvalues of $U$ must come in conjugate pairs, while the real eigenvalues can only take the values $+1$ or $-1$.
 
\begin{lemma}
The number of non-accidental $\pm 1$ eigenvalues of $U$ equals $|\nu\pm\nu_\star|$.
\label{Invertilemma} \end{lemma}
 
If we define the projection operators $Q^\pm$ and $Q_\star^\pm$ as
\beq
Q^\pm = \frac{1\pm\epsilon}{2}\ ,\quad
Q_\star^\pm = \frac{1\pm\epsilon_\star}{2}\ ,
\eeq
and 
 \beq
 \Delta^\pm\equiv (1 \pm U)\ .
 \eeq
then they are related by
 \beq
\Delta^+ &=& \epsilon (\epsilon+\epsilon_\star) =2 \epsilon\left(Q^+ - Q^-_\star\right) =2 \epsilon\left(Q^+_\star - Q^-\right)\ ,\cr
\Delta^- &=& \epsilon (\epsilon-\epsilon_\star) =2 \epsilon\left(Q^+ - Q^+_\star\right) = 2\epsilon\left(Q^-_\star - Q^-\right)\ .
\eeq
 If these are all $2N\times 2N$ matrices, then $\epsilon$ is rank $2N$ while $Q^\pm $ and $ Q^\pm_\star$ are rank $N\pm\nu$ and $N\pm\nu_\star$ respectively.   It follows from the subadditivity of rank that 
\beq
\text{rank}\,\Delta^\pm \le  2N -| \nu\mp \nu_\star|\ ,
\eeq
meaning that $\Delta^\pm$ must have at least $| \nu\mp\nu_\star|$ zero eigenvalues. We will assume that typically this inequality is saturated, and that additional accidental zeromodes are not important.  Thus $U$ has $| \nu\pm\nu_\star|$ non-accidental eigenvalues equal to $\pm 1$ respectively.  It immediately follows that $(1+U)$ is not invertible when $\nu\ne\nu_\star$.

\begin{lemma}
Eigenstates of $U$ corresponding to eigenvalues $\pm 1$ can be taken to be simultaneous eigenstates of $\epsilon$ and $\epsilon_\star$.
 \end{lemma}
From \eq{conj}, if $\ket{+1}$ satisfies $U\ket{+1}=\ket{+1}$, then so does the state $\epsilon\ket{+1}$.  Therefore we can take  as our basis vectors the nonvanishing $Q^\pm\ket{+1}$ vectors.  These are eigenstates of $\epsilon$ with eigenvalues $\pm1$ respectively.  It follows from the definition $U=\epsilon\epsilon_\star$ that these states are also simultaneous eigenstates of $\epsilon_\star$ with eigenvalue $\pm 1$ respectively.

Similar reasoning for an eigenstate $\ket{-1}$ satisfying $U\ket{-1}=-\ket{-1}$ leads to states $Q^\pm\ket{-1}$ which are eigenstates of $\epsilon$ with eigenvalue $\pm1$, and of $\epsilon_\star$ with eigenvalue $\mp 1$ respectively.

\begin{lemma}
The non-accidental eigenstates of $U$ with eigenvalue $\pm1$ are simultaneous eigenstates of $\epsilon$ with eigenvalue $ {\rm Sign}[\nu\pm \nu_\star]$
 \end{lemma}
 
An index theorem for $U$ can be proven by adapting the methods of \cite{Luscher:1998pqa}. We note that $\Delta^\pm$ satisfy  equations similar to the GW equation, with $\epsilon$ playing the role of $\gamma_5$:
\beq
\left\{ \Delta^\pm,\epsilon\,\right\} = \Delta^\pm \,\epsilon\,\Delta^\pm\ ,
\eqn{GWprime}
\eeq  
from which one can derive the relation
\beq
(z-\Delta^\pm)\epsilon\,(z-\Delta^\pm)\hspace{1.95 in}\  \cr &&\cr && \hspace{-3.2in}\ = z(2-z)\epsilon - (1-z)\left[(z-\Delta^\pm) \epsilon + \epsilon\,(z-\Delta^\pm)\right]
\eeq
for arbitrary complex variable $z$. Multiplying this on the right by $(z-\Delta^\pm)^{-1}$ and taking the trace yields
\beq
\hspace{-.3in}-\Tr\epsilon\,\Delta^\pm  &=& -(2-z)\Tr\epsilon \nonumber \\
&+& z(2-z)\Tr \epsilon\,(z-\Delta^\pm)^{-1}\ .
\eqn{indeq}\eeq
Integrating this equation on a sufficiently small circle about the origin in the complex $z$ plane yields the index theorem. The integral 
\beq
(\nu\mp\nu_\star) =\Tr\epsilon\, \CP^\pm_0 =  (\gothn^\pm_+ - \gothn^\pm_-)\ ,
\eqn{epsind}\eeq
where 
\beq
\CP^\pm_0 = \oint \frac{dz}{2\pi i} (z-\Delta^\pm)^{-1}
\eeq
 serves as a projector onto the space of zeromodes of $\Delta^\pm$.  In \eq{epsind},   $\gothn^\pm_+$ is the number of states in the kernel of $(1\pm U)$   that are simultaneously eigenstates of $\epsilon$ with eigenvalue  $+1$,  and $\gothn^\pm_-$ is the number of states in the kernel of $(1\pm U)$    that are simultaneously eigenstates of $\epsilon$ with eigenvalue  $-1$.  Again assuming that there are no accidental zeromodes, we conclude that 
 \beq
 \{\gothn^+_+,\gothn^+_-\} = \begin{cases}
  \{ |\nu-\nu_\star|,0\} & \nu>\nu_\star\\
  \{0,0\} & \nu=\nu_\star\\
  \{0,  |\nu-\nu_\star|\} & \nu<\nu_\star
  \end{cases}\ ,
  \eeq
  and
  \beq
 \{\gothn^-_+,\gothn^-_-\} = \begin{cases}
  \{ |\nu+\nu_\star|,0\} & \nu>-\nu_\star\\
  \{0,0\} & \nu=-\nu_\star\\
  \{0,  |\nu+\nu_\star|\} & \nu<-\nu_\star
  \end{cases}\ .
  \eeq
  Accidental zeromodes will occur when a pair of states with eigenvalues $e^{\pm i \delta}$ exhibits a level crossing with $\delta$ passing through a multiple of $\pi$.  For the particular gauge field for which such a crossing occurs, both $\gothn^-_\pm$ will change by one if at $\delta=0\ \text{Mod}[2\pi]$, while both $\gothn^+_\pm$ will change by one if at $\delta=\pi\ \text{Mod}[2\pi]$.

\section{Matrix elements of $\hat\CE_\chi$ for general gauge field topology}

Our proposal for $\hat\CD_\chi$ in \eq{ChiralOverOp} is 
\beq
\hat\CD_\chi &=& 1 + \gamma_5 \hat\CE_\chi\ ,\cr
\hat\CE_\chi &=& 1 - (1-\epsilon_\star)\frac{1}{1+U}(1-\epsilon)\ .
\eeq
where $U=\epsilon\epsilon_\star$ as in \eq{Udef}. In Sec.~\ref{gweq} we showed that $\hat\CD_\chi$ obeyed the GW equation whenever the matrix $\Delta^+=(1+U)$ is invertible. This only occurs when $\nu=\nu_\star$, as was shown in the previous appendix, Lemma~\ref{Invertilemma}. Using the definition of these quantities in \eq{nudef}, we conclude that $1+U$ is only invertible  when the gradient flow preserves topology.  In order to show that $\hat\CD_\chi$ generally obeys the GW equation for any type of gauge-covariant flow equation, one must generalize $\hat\CE_\chi$ to the case when $\Delta^+=(1+U)$ is not invertible, which we do here.   This will also allow us  to compute $\Tr\hat\CE_\chi$, which is the chiral  index of $\hat\CD_\chi$ and yields the chiral anomaly.    

To better understand $\hat\CE_\chi$ when  $\nu\ne\nu_\star$ we can use eigenstates of $U$ to define the Hilbert space, which is divided into three distinct subspaces spanned by states $\ket{I}$, $\ket{m}$, $\ket{i}$ where
\begin{equation}
\begin{aligned}
&U\ket{I} &=& \ket{I}\ ,\  &I &= 1,\ldots,|\nu+\nu_\star| \cr
&U\ket{i} &= &-\ket{i}\ ,\ & i &= 1,\ldots,|\nu-\nu_\star| \cr
&U\ket{m} &=& \eta_m\ket{m}\ ,\  &m &= 1,\ldots,2N-|\nu+\nu_\star|-|\nu-\nu_\star| \ ,
\end{aligned}
\eqn{Uaction}\end{equation}
with $|\eta_m|=1$ and $\eta_m\ne \pm 1$; the complex eigenvalues come in complex conjugate pairs. The action of $\epsilon$ and $\epsilon_\star$ on these states is
\begin{equation}
\begin{aligned}
&\epsilon \ket{I} &=&+\epsilon_\star \ket{I}   &=&\text{ sign}[\nu+\nu_\star] \ket{I} \cr
&\epsilon\ket{i} &=&-\epsilon_\star \ket{i}   &=&\text{ sign}[\nu-\nu_\star] \ket{i} \ ,
\end{aligned}
\eqn{eaction1}\end{equation}
while
\beq
\epsilon \ket{m} \equiv \ket{\mybar m}\ ,\quad \epsilon_\star \ket{m} \equiv \eta_m\ket{\mybar m}
\eqn{eaction2}\eeq
where $\ket{\mybar m}$ is the state with $U$ eigenvalue $\eta_m^*$, and we have made a particular phase choice in its definition.

It is evident that in this basis the matrix $\hat\CE_\chi $ is diagonal in the $\ket{I}$, $\ket{i}$ sectors, and is block diagonal in the remaining space, in $2\times 2$ blocks acting on each $\left\{\ket{m}, \ket{\bar m}\right\}$ pair of states. Using the results in the above equations, the nonzero matrix elements of $\hat\CE_\chi$ are given by
\beq
\expect{I}{\hat\CE_\chi}{I} &=& \text{ sign}[\nu+\nu_\star] \ ,\cr && \cr
\expect{i}{\hat\CE_\chi}{i} &=& \xi_i\ ,\cr && \cr
\begin{pmatrix}\bra{m}\\ \bra{\mybar m}\end{pmatrix}\,\hat\CE_\chi\,\begin{pmatrix}\ket{m}&\ket{\mybar m}\end{pmatrix} &=&\frac{1}{1+\eta_m}
\begin{pmatrix} \eta_m -1 & 2\\ 2\eta_m & 1-\eta_m \end{pmatrix}\cr && \ .
\eqn{ME}\eeq
In order for $\hat\CD_\chi$ to satisfy the GW equation, $\hat\CE_\chi^2=1$ and, given that
\beq
\left( \text{ sign}[\nu+\nu_\star] \right)^2 &=& 1\ ,\cr &&\cr
 \frac{1}{(1+\eta_m)^2}
\begin{pmatrix} \eta_m -1 & 2\\ 2\eta_m & 1-\eta_m \end{pmatrix} ^2 &=& \begin{pmatrix}1 & 0 \\ 0 & 1\end{pmatrix}\ ,
\eeq
it follows that $\xi_i^2 =  1$ for every index $i$.

If we naively try to compute the $\xi_i = \expect{i}{\hat\CE_\chi}{i}$ matrix elements from \eq{Uaction} and \eq{eaction1} we will find the ratio of two vanishing quantities. Nevertheless, as was shown in Sec.~\ref{chiprops} the eigenvalues of $\hat\CE_\chi$ are bounded between $\pm1$, and as the $\xi_i$ are such eigenvalues, it follows that
\beq
-1\le\xi_i\le 1\ .
\eeq
Furthermore, we argued that except at exceptional gauge fields that relate the spectrum of $H$ to that of $H_\star$ in a special way, the $\xi_i$ must saturate the bounds with $\xi_i=\pm 1$.

What this argument does not tell us is how many of the $\xi_i$ equal $+1$ and how many equal $-1$; it is not possible to determine this knowing $\nu$ and $\nu_\star$ alone.  To see this, consider rescaling $H$ and $H_\star$ by  number $c$ and $c_\star$ respectively. These rescalings do not affect $\epsilon$ or $ \epsilon_\star$.  If $c\to 0$ while $c_\star=1$ then we can ignore $T $ in \eq{CEdefC} and  $\hat\CE_\chi \to  \epsilon$; if we do the reverse, then we get $\hat\CE_\chi \to \epsilon_\star$.   Thus by changing $H$ and $H_\star$ relative to each other without changing the signs of their eigenvalues, we can change the trace of $\hat\CE_\chi$, presumably  by jumps of 2 as successive $\xi_i$ change sign.  

We conclude that, except for exceptional gauge fields, $\hat\CE_\chi$ obeys the GW equation, and that it is precisely when one passes through  those special gauge fields that  it is possible for the trace of $\hat\CE_\chi$ to jump by $\pm2$.

For the anomaly we need to compute the trace of $\hat\CE_\chi$. Using the above analysis, we find
\beq
\Tr\hat\CE_\chi = (\nu+\nu_\star)+ \sum_{i=1}^{|\nu-\nu_\star|} \xi_i 
\eqn{TrEX}
\eeq
where the first term comes from the $\ket{I}$ space where $\epsilon\epsilon_* = +1$,  and the second term arises from the $\ket{i}$ space where $\epsilon\epsilon_* = -1$.  Note that the complex eigenvalue sectors do not contribute as the corresponding $2\times 2$ matrix elements in \eq{ME} are traceless.

 \end{appendix}
\bibliography{CGT.bib}

\begin{thebibliography}{55}%
\makeatletter
\providecommand \@ifxundefined [1]{%
 \@ifx{#1\undefined}
}%
\providecommand \@ifnum [1]{%
 \ifnum #1\expandafter \@firstoftwo
 \else \expandafter \@secondoftwo
 \fi
}%
\providecommand \@ifx [1]{%
 \ifx #1\expandafter \@firstoftwo
 \else \expandafter \@secondoftwo
 \fi
}%
\providecommand \natexlab [1]{#1}%
\providecommand \enquote  [1]{``#1''}%
\providecommand \bibnamefont  [1]{#1}%
\providecommand \bibfnamefont [1]{#1}%
\providecommand \citenamefont [1]{#1}%
\providecommand \href@noop [0]{\@secondoftwo}%
\providecommand \href [0]{\begingroup \@sanitize@url \@href}%
\providecommand \@href[1]{\@@startlink{#1}\@@href}%
\providecommand \@@href[1]{\endgroup#1\@@endlink}%
\providecommand \@sanitize@url [0]{\catcode `\\12\catcode `\$12\catcode
  `\&12\catcode `\#12\catcode `\^12\catcode `\_12\catcode `\%12\relax}%
\providecommand \@@startlink[1]{}%
\providecommand \@@endlink[0]{}%
\providecommand \url  [0]{\begingroup\@sanitize@url \@url }%
\providecommand \@url [1]{\endgroup\@href {#1}{\urlprefix }}%
\providecommand \urlprefix  [0]{URL }%
\providecommand \Eprint [0]{\href }%
\providecommand \doibase [0]{http://dx.doi.org/}%
\providecommand \selectlanguage [0]{\@gobble}%
\providecommand \bibinfo  [0]{\@secondoftwo}%
\providecommand \bibfield  [0]{\@secondoftwo}%
\providecommand \translation [1]{[#1]}%
\providecommand \BibitemOpen [0]{}%
\providecommand \bibitemStop [0]{}%
\providecommand \bibitemNoStop [0]{.\EOS\space}%
\providecommand \EOS [0]{\spacefactor3000\relax}%
\providecommand \BibitemShut  [1]{\csname bibitem#1\endcsname}%
\let\auto@bib@innerbib\@empty
%</preamble>
\bibitem [{\citenamefont {Alvarez-Gaume}\ and\ \citenamefont
  {Ginsparg}(1984)}]{AlvarezGaume:1983cs}%
  \BibitemOpen
  \bibfield  {author} {\bibinfo {author} {\bibfnamefont {L.}~\bibnamefont
  {Alvarez-Gaume}}\ and\ \bibinfo {author} {\bibfnamefont {P.~H.}\ \bibnamefont
  {Ginsparg}},\ }\href {\doibase 10.1016/0550-3213(84)90487-5} {\bibfield
  {journal} {\bibinfo  {journal} {Nucl. Phys.}\ }\textbf {\bibinfo {volume}
  {B243}},\ \bibinfo {pages} {449} (\bibinfo {year} {1984})}\BibitemShut
  {NoStop}%
%%CITATION = NUPHA,B243,449;%%
\bibitem [{\citenamefont {Alvarez-Gaume}\ \emph {et~al.}(1986)\citenamefont
  {Alvarez-Gaume}, \citenamefont {Della~Pietra},\ and\ \citenamefont
  {Della~Pietra}}]{AlvarezGaume:1985di}%
  \BibitemOpen
  \bibfield  {author} {\bibinfo {author} {\bibfnamefont {L.}~\bibnamefont
  {Alvarez-Gaume}}, \bibinfo {author} {\bibfnamefont {S.}~\bibnamefont
  {Della~Pietra}}, \ and\ \bibinfo {author} {\bibfnamefont {V.}~\bibnamefont
  {Della~Pietra}},\ }\href {\doibase 10.1016/0370-2693(86)91373-0} {\bibfield
  {journal} {\bibinfo  {journal} {Phys. Lett.}\ }\textbf {\bibinfo {volume}
  {B166}},\ \bibinfo {pages} {177} (\bibinfo {year} {1986})}\BibitemShut
  {NoStop}%
%%CITATION = PHLTA,B166,177;%%
\bibitem [{\citenamefont {Fujikawa}(1979)}]{Fujikawa:1979ay}%
  \BibitemOpen
  \bibfield  {author} {\bibinfo {author} {\bibfnamefont {K.}~\bibnamefont
  {Fujikawa}},\ }\href {\doibase 10.1103/PhysRevLett.42.1195} {\bibfield
  {journal} {\bibinfo  {journal} {Phys. Rev. Lett.}\ }\textbf {\bibinfo
  {volume} {42}},\ \bibinfo {pages} {1195} (\bibinfo {year}
  {1979})}\BibitemShut {NoStop}%
%%CITATION = PRLTA,42,1195;%%
\bibitem [{\citenamefont {Adler}(1969)}]{Adler:1969gk}%
  \BibitemOpen
  \bibfield  {author} {\bibinfo {author} {\bibfnamefont {S.~L.}\ \bibnamefont
  {Adler}},\ }\href {\doibase 10.1103/PhysRev.177.2426} {\bibfield  {journal}
  {\bibinfo  {journal} {Phys. Rev.}\ }\textbf {\bibinfo {volume} {177}},\
  \bibinfo {pages} {2426} (\bibinfo {year} {1969})}\BibitemShut {NoStop}%
%%CITATION = PHRVA,177,2426;%%
\bibitem [{\citenamefont {{Bell}}\ and\ \citenamefont
  {{Jackiw}}(1969)}]{Bell:1969ab}%
  \BibitemOpen
  \bibfield  {author} {\bibinfo {author} {\bibfnamefont {J.~S.}\ \bibnamefont
  {{Bell}}}\ and\ \bibinfo {author} {\bibfnamefont {R.}~\bibnamefont
  {{Jackiw}}},\ }\href {\doibase 10.1007/BF02823296} {\bibfield  {journal}
  {\bibinfo  {journal} {Nuovo Cimento A Serie}\ }\textbf {\bibinfo {volume}
  {60}},\ \bibinfo {pages} {47} (\bibinfo {year} {1969})}\BibitemShut {NoStop}%
\bibitem [{\citenamefont {Ginsparg}\ and\ \citenamefont
  {Wilson}(1982)}]{Ginsparg:1981bj}%
  \BibitemOpen
  \bibfield  {author} {\bibinfo {author} {\bibfnamefont {P.~H.}\ \bibnamefont
  {Ginsparg}}\ and\ \bibinfo {author} {\bibfnamefont {K.~G.}\ \bibnamefont
  {Wilson}},\ }\href {\doibase 10.1103/PhysRevD.25.2649} {\bibfield  {journal}
  {\bibinfo  {journal} {Phys. Rev.}\ }\textbf {\bibinfo {volume} {D25}},\
  \bibinfo {pages} {2649} (\bibinfo {year} {1982})}\BibitemShut {NoStop}%
%%CITATION = PHRVA,D25,2649;%%
\bibitem [{\citenamefont {Hasenfratz}\ \emph {et~al.}(1998)\citenamefont
  {Hasenfratz}, \citenamefont {Laliena},\ and\ \citenamefont
  {Niedermayer}}]{Hasenfratz:1998ri}%
  \BibitemOpen
  \bibfield  {author} {\bibinfo {author} {\bibfnamefont {P.}~\bibnamefont
  {Hasenfratz}}, \bibinfo {author} {\bibfnamefont {V.}~\bibnamefont {Laliena}},
  \ and\ \bibinfo {author} {\bibfnamefont {F.}~\bibnamefont {Niedermayer}},\
  }\href {\doibase 10.1016/S0370-2693(98)00315-3} {\bibfield  {journal}
  {\bibinfo  {journal} {Phys. Lett.}\ }\textbf {\bibinfo {volume} {B427}},\
  \bibinfo {pages} {125} (\bibinfo {year} {1998})},\ \Eprint
  {http://arxiv.org/abs/hep-lat/9801021} {arXiv:hep-lat/9801021 [hep-lat]}
  \BibitemShut {NoStop}%
%%CITATION = HEP-LAT/9801021;%%
\bibitem [{\citenamefont {Luscher}(1998)}]{Luscher:1998pqa}%
  \BibitemOpen
  \bibfield  {author} {\bibinfo {author} {\bibfnamefont {M.}~\bibnamefont
  {Luscher}},\ }\href {\doibase 10.1016/S0370-2693(98)00423-7} {\bibfield
  {journal} {\bibinfo  {journal} {Phys. Lett.}\ }\textbf {\bibinfo {volume}
  {B428}},\ \bibinfo {pages} {342} (\bibinfo {year} {1998})},\ \Eprint
  {http://arxiv.org/abs/hep-lat/9802011} {arXiv:hep-lat/9802011 [hep-lat]}
  \BibitemShut {NoStop}%
%%CITATION = HEP-LAT/9802011;%%
\bibitem [{\citenamefont {Narayanan}\ and\ \citenamefont
  {Neuberger}(1994)}]{Narayanan:1993sk}%
  \BibitemOpen
  \bibfield  {author} {\bibinfo {author} {\bibfnamefont {R.}~\bibnamefont
  {Narayanan}}\ and\ \bibinfo {author} {\bibfnamefont {H.}~\bibnamefont
  {Neuberger}},\ }\href {\doibase 10.1016/0550-3213(94)90393-X} {\bibfield
  {journal} {\bibinfo  {journal} {Nucl. Phys.}\ }\textbf {\bibinfo {volume}
  {B412}},\ \bibinfo {pages} {574} (\bibinfo {year} {1994})},\ \Eprint
  {http://arxiv.org/abs/hep-lat/9307006} {arXiv:hep-lat/9307006 [hep-lat]}
  \BibitemShut {NoStop}%
%%CITATION = HEP-LAT/9307006;%%
\bibitem [{\citenamefont {Narayanan}\ and\ \citenamefont
  {Neuberger}(1995)}]{Narayanan:1994gw}%
  \BibitemOpen
  \bibfield  {author} {\bibinfo {author} {\bibfnamefont {R.}~\bibnamefont
  {Narayanan}}\ and\ \bibinfo {author} {\bibfnamefont {H.}~\bibnamefont
  {Neuberger}},\ }\href {\doibase 10.1016/0550-3213(95)00111-5} {\bibfield
  {journal} {\bibinfo  {journal} {Nucl. Phys.}\ }\textbf {\bibinfo {volume}
  {B443}},\ \bibinfo {pages} {305} (\bibinfo {year} {1995})},\ \Eprint
  {http://arxiv.org/abs/hep-th/9411108} {arXiv:hep-th/9411108 [hep-th]}
  \BibitemShut {NoStop}%
%%CITATION = HEP-TH/9411108;%%
\bibitem [{\citenamefont {Neuberger}(1998{\natexlab{a}})}]{Neuberger:1997fp}%
  \BibitemOpen
  \bibfield  {author} {\bibinfo {author} {\bibfnamefont {H.}~\bibnamefont
  {Neuberger}},\ }\href {\doibase 10.1016/S0370-2693(97)01368-3} {\bibfield
  {journal} {\bibinfo  {journal} {Phys.Lett.}\ }\textbf {\bibinfo {volume}
  {B417}},\ \bibinfo {pages} {141} (\bibinfo {year} {1998}{\natexlab{a}})},\
  \Eprint {http://arxiv.org/abs/hep-lat/9707022} {arXiv:hep-lat/9707022
  [hep-lat]} \BibitemShut {NoStop}%
%%CITATION = HEP-LAT/9707022;%%
\bibitem [{\citenamefont {Neuberger}(1998{\natexlab{b}})}]{Neuberger:1998wv}%
  \BibitemOpen
  \bibfield  {author} {\bibinfo {author} {\bibfnamefont {H.}~\bibnamefont
  {Neuberger}},\ }\href {\doibase 10.1016/S0370-2693(98)00355-4} {\bibfield
  {journal} {\bibinfo  {journal} {Phys. Lett.}\ }\textbf {\bibinfo {volume}
  {B427}},\ \bibinfo {pages} {353} (\bibinfo {year} {1998}{\natexlab{b}})},\
  \Eprint {http://arxiv.org/abs/hep-lat/9801031} {arXiv:hep-lat/9801031
  [hep-lat]} \BibitemShut {NoStop}%
%%CITATION = HEP-LAT/9801031;%%
\bibitem [{\citenamefont {Golterman}\ and\ \citenamefont
  {Shamir}(2004)}]{golterman2004s}%
  \BibitemOpen
  \bibfield  {author} {\bibinfo {author} {\bibfnamefont {M.}~\bibnamefont
  {Golterman}}\ and\ \bibinfo {author} {\bibfnamefont {Y.}~\bibnamefont
  {Shamir}},\ }\href@noop {} {\bibfield  {journal} {\bibinfo  {journal}
  {Physical Review D}\ }\textbf {\bibinfo {volume} {70}},\ \bibinfo {pages}
  {094506} (\bibinfo {year} {2004})}\BibitemShut {NoStop}%
\bibitem [{\citenamefont {Golterman}(2001)}]{Golterman:2000hr}%
  \BibitemOpen
  \bibfield  {author} {\bibinfo {author} {\bibfnamefont {M.}~\bibnamefont
  {Golterman}},\ }\href {\doibase 10.1016/S0920-5632(01)00953-7} {\bibfield
  {journal} {\bibinfo  {journal} {Nucl. Phys. Proc. Suppl.}\ }\textbf {\bibinfo
  {volume} {94}},\ \bibinfo {pages} {189} (\bibinfo {year} {2001})},\ \Eprint
  {http://arxiv.org/abs/hep-lat/0011027} {arXiv:hep-lat/0011027 [hep-lat]}
  \BibitemShut {NoStop}%
%%CITATION = HEP-LAT/0011027;%%
\bibitem [{\citenamefont {Luscher}(2002)}]{Luscher:2000hn}%
  \BibitemOpen
  \bibfield  {author} {\bibinfo {author} {\bibfnamefont {M.}~\bibnamefont
  {Luscher}},\ }\bibfield  {booktitle} {\emph {\bibinfo {booktitle} {{Theory
  and experiment heading for new physics. Proceedings, 38th course of the
  International School of subnuclear physics, Erice, Italy, August 27-September
  5, 2000}}},\ }\href {\doibase 10.1142/9789812778253_0002} {\bibfield
  {journal} {\bibinfo  {journal} {Subnucl. Ser.}\ }\textbf {\bibinfo {volume}
  {38}},\ \bibinfo {pages} {41} (\bibinfo {year} {2002})},\ \Eprint
  {http://arxiv.org/abs/hep-th/0102028} {arXiv:hep-th/0102028 [hep-th]}
  \BibitemShut {NoStop}%
%%CITATION = HEP-TH/0102028;%%
\bibitem [{\citenamefont {Wen}(2013)}]{Wen:2013ppa}%
  \BibitemOpen
  \bibfield  {author} {\bibinfo {author} {\bibfnamefont {X.-G.}\ \bibnamefont
  {Wen}},\ }\href {\doibase 10.1088/0256-307X/30/11/111101} {\bibfield
  {journal} {\bibinfo  {journal} {Chin. Phys. Lett.}\ }\textbf {\bibinfo
  {volume} {30}},\ \bibinfo {pages} {111101} (\bibinfo {year} {2013})},\
  \Eprint {http://arxiv.org/abs/1305.1045} {arXiv:1305.1045 [hep-lat]}
  \BibitemShut {NoStop}%
%%CITATION = ARXIV:1305.1045;%%
\bibitem [{\citenamefont {Giedt}\ \emph {et~al.}(2014)\citenamefont {Giedt},
  \citenamefont {Chen},\ and\ \citenamefont {Poppitz}}]{Giedt:2014pha}%
  \BibitemOpen
  \bibfield  {author} {\bibinfo {author} {\bibfnamefont {J.}~\bibnamefont
  {Giedt}}, \bibinfo {author} {\bibfnamefont {C.}~\bibnamefont {Chen}}, \ and\
  \bibinfo {author} {\bibfnamefont {E.}~\bibnamefont {Poppitz}},\ }\bibfield
  {booktitle} {\emph {\bibinfo {booktitle} {{Proceedings, 31st International
  Symposium on Lattice Field Theory (Lattice 2013)}}},\ }\href@noop {}
  {\bibfield  {journal} {\bibinfo  {journal} {PoS}\ }\textbf {\bibinfo {volume}
  {LATTICE2013}},\ \bibinfo {pages} {131} (\bibinfo {year} {2014})},\ \Eprint
  {http://arxiv.org/abs/1403.5146} {arXiv:1403.5146 [hep-lat]} \BibitemShut
  {NoStop}%
%%CITATION = ARXIV:1403.5146;%%
\bibitem [{\citenamefont {You}\ and\ \citenamefont
  {Xu}(2015)}]{you2015interacting}%
  \BibitemOpen
  \bibfield  {author} {\bibinfo {author} {\bibfnamefont {Y.-Z.}\ \bibnamefont
  {You}}\ and\ \bibinfo {author} {\bibfnamefont {C.}~\bibnamefont {Xu}},\
  }\href@noop {} {\bibfield  {journal} {\bibinfo  {journal} {Physical Review
  B}\ }\textbf {\bibinfo {volume} {91}},\ \bibinfo {pages} {125147} (\bibinfo
  {year} {2015})}\BibitemShut {NoStop}%
\bibitem [{\citenamefont {Grabowska}\ and\ \citenamefont
  {Kaplan}(2016)}]{Grabowska:2015qpk}%
  \BibitemOpen
  \bibfield  {author} {\bibinfo {author} {\bibfnamefont {D.~M.}\ \bibnamefont
  {Grabowska}}\ and\ \bibinfo {author} {\bibfnamefont {D.~B.}\ \bibnamefont
  {Kaplan}},\ }\href {\doibase 10.1103/PhysRevLett.116.211602} {\bibfield
  {journal} {\bibinfo  {journal} {Phys. Rev. Lett.}\ }\textbf {\bibinfo
  {volume} {116}},\ \bibinfo {pages} {211602} (\bibinfo {year} {2016})},\
  \Eprint {http://arxiv.org/abs/1511.03649} {arXiv:1511.03649 [hep-lat]}
  \BibitemShut {NoStop}%
%%CITATION = ARXIV:1511.03649;%%
\bibitem [{\citenamefont {Grabowska}()}]{Grabowska:Lattice}%
  \BibitemOpen
  \bibfield  {author} {\bibinfo {author} {\bibfnamefont {D.~M.}\ \bibnamefont
  {Grabowska}},\ }\href@noop {} {\bibinfo  {journal} {Talk delivered at Lattice
  2016:
  \url{https://conference.ippp.dur.ac.uk/event/470/session/16/contribution/364}}\
  }\BibitemShut {NoStop}%
\bibitem [{\citenamefont {Kaplan}()}]{Kaplan:Lattice}%
  \BibitemOpen
\bibfield  {journal} {  }\bibfield  {author} {\bibinfo {author} {\bibfnamefont
  {D.~B.}\ \bibnamefont {Kaplan}},\ }\href@noop {} {\bibinfo  {journal} {Talk
  delivered at Lattice 2016:
  \url{https://conference.ippp.dur.ac.uk/event/470/session/1/contribution/398}}\
  }\BibitemShut {NoStop}%
\bibitem [{\citenamefont {Kaplan}(1992)}]{Kaplan:1992bt}%
  \BibitemOpen
\bibfield  {journal} {  }\bibfield  {author} {\bibinfo {author} {\bibfnamefont
  {D.~B.}\ \bibnamefont {Kaplan}},\ }\href {\doibase
  10.1016/0370-2693(92)91112-M} {\bibfield  {journal} {\bibinfo  {journal}
  {Phys. Lett.}\ }\textbf {\bibinfo {volume} {B288}},\ \bibinfo {pages} {342}
  (\bibinfo {year} {1992})},\ \Eprint {http://arxiv.org/abs/hep-lat/9206013}
  {arXiv:hep-lat/9206013 [hep-lat]} \BibitemShut {NoStop}%
%%CITATION = HEP-LAT/9206013;%%
\bibitem [{\citenamefont {Kaplan}(2009)}]{Kaplan:2009yg}%
  \BibitemOpen
  \bibfield  {author} {\bibinfo {author} {\bibfnamefont {D.~B.}\ \bibnamefont
  {Kaplan}},\ }in\ \href
  {http://inspirehep.net/record/839996/files/arXiv:0912.2560.pdf} {\emph
  {\bibinfo {booktitle} {{Modern perspectives in lattice QCD: Quantum field
  theory and high performance computing. Proceedings, International School,
  93rd Session, Les Houches, France, August 3-28, 2009}}}}\ (\bibinfo {year}
  {2009})\ pp.\ \bibinfo {pages} {223--272},\ \Eprint
  {http://arxiv.org/abs/0912.2560} {arXiv:0912.2560 [hep-lat]} \BibitemShut
  {NoStop}%
%%CITATION = ARXIV:0912.2560;%%
\bibitem [{\citenamefont {Callan}\ and\ \citenamefont
  {Harvey}(1985)}]{Callan:1984sa}%
  \BibitemOpen
  \bibfield  {author} {\bibinfo {author} {\bibfnamefont {C.~G.}\ \bibnamefont
  {Callan}, \bibfnamefont {Jr.}}\ and\ \bibinfo {author} {\bibfnamefont
  {J.~A.}\ \bibnamefont {Harvey}},\ }\href {\doibase
  10.1016/0550-3213(85)90489-4} {\bibfield  {journal} {\bibinfo  {journal}
  {Nucl. Phys.}\ }\textbf {\bibinfo {volume} {B250}},\ \bibinfo {pages} {427}
  (\bibinfo {year} {1985})}\BibitemShut {NoStop}%
%%CITATION = NUPHA,B250,427;%%
\bibitem [{\citenamefont {Golterman}\ \emph {et~al.}(1993)\citenamefont
  {Golterman}, \citenamefont {Jansen},\ and\ \citenamefont
  {Kaplan}}]{Golterman:1992ub}%
  \BibitemOpen
  \bibfield  {author} {\bibinfo {author} {\bibfnamefont {M.~F.~L.}\
  \bibnamefont {Golterman}}, \bibinfo {author} {\bibfnamefont {K.}~\bibnamefont
  {Jansen}}, \ and\ \bibinfo {author} {\bibfnamefont {D.~B.}\ \bibnamefont
  {Kaplan}},\ }\href {\doibase 10.1016/0370-2693(93)90692-B} {\bibfield
  {journal} {\bibinfo  {journal} {Phys. Lett.}\ }\textbf {\bibinfo {volume}
  {B301}},\ \bibinfo {pages} {219} (\bibinfo {year} {1993})},\ \Eprint
  {http://arxiv.org/abs/hep-lat/9209003} {arXiv:hep-lat/9209003 [hep-lat]}
  \BibitemShut {NoStop}%
%%CITATION = HEP-LAT/9209003;%%
\bibitem [{\citenamefont {Jansen}\ and\ \citenamefont
  {Schmaltz}(1992)}]{Jansen:1992tw}%
  \BibitemOpen
  \bibfield  {author} {\bibinfo {author} {\bibfnamefont {K.}~\bibnamefont
  {Jansen}}\ and\ \bibinfo {author} {\bibfnamefont {M.}~\bibnamefont
  {Schmaltz}},\ }\href {\doibase 10.1016/0370-2693(92)91335-7} {\bibfield
  {journal} {\bibinfo  {journal} {Phys. Lett.}\ }\textbf {\bibinfo {volume}
  {B296}},\ \bibinfo {pages} {374} (\bibinfo {year} {1992})},\ \Eprint
  {http://arxiv.org/abs/hep-lat/9209002} {arXiv:hep-lat/9209002 [hep-lat]}
  \BibitemShut {NoStop}%
%%CITATION = HEP-LAT/9209002;%%
\bibitem [{\citenamefont {Shamir}(1993)}]{Shamir:1993zy}%
  \BibitemOpen
  \bibfield  {author} {\bibinfo {author} {\bibfnamefont {Y.}~\bibnamefont
  {Shamir}},\ }\href {\doibase 10.1016/0550-3213(93)90162-I} {\bibfield
  {journal} {\bibinfo  {journal} {Nucl. Phys.}\ }\textbf {\bibinfo {volume}
  {B406}},\ \bibinfo {pages} {90} (\bibinfo {year} {1993})},\ \Eprint
  {http://arxiv.org/abs/hep-lat/9303005} {arXiv:hep-lat/9303005 [hep-lat]}
  \BibitemShut {NoStop}%
%%CITATION = HEP-LAT/9303005;%%
\bibitem [{\citenamefont {Furman}\ and\ \citenamefont
  {Shamir}(1995)}]{Furman:1994ky}%
  \BibitemOpen
  \bibfield  {author} {\bibinfo {author} {\bibfnamefont {V.}~\bibnamefont
  {Furman}}\ and\ \bibinfo {author} {\bibfnamefont {Y.}~\bibnamefont
  {Shamir}},\ }\href {\doibase 10.1016/0550-3213(95)00031-M} {\bibfield
  {journal} {\bibinfo  {journal} {Nucl. Phys.}\ }\textbf {\bibinfo {volume}
  {B439}},\ \bibinfo {pages} {54} (\bibinfo {year} {1995})},\ \Eprint
  {http://arxiv.org/abs/hep-lat/9405004} {arXiv:hep-lat/9405004 [hep-lat]}
  \BibitemShut {NoStop}%
%%CITATION = HEP-LAT/9405004;%%
\bibitem [{\citenamefont {Jansen}(1992)}]{Jansen:1992yj}%
  \BibitemOpen
  \bibfield  {author} {\bibinfo {author} {\bibfnamefont {K.}~\bibnamefont
  {Jansen}},\ }\href {\doibase 10.1016/0370-2693(92)91113-N} {\bibfield
  {journal} {\bibinfo  {journal} {Phys. Lett.}\ }\textbf {\bibinfo {volume}
  {B288}},\ \bibinfo {pages} {348} (\bibinfo {year} {1992})},\ \Eprint
  {http://arxiv.org/abs/hep-lat/9206014} {arXiv:hep-lat/9206014 [hep-lat]}
  \BibitemShut {NoStop}%
%%CITATION = HEP-LAT/9206014;%%
\bibitem [{\citenamefont {Kaplan}(1993)}]{Kaplan:1992sg}%
  \BibitemOpen
  \bibfield  {author} {\bibinfo {author} {\bibfnamefont {D.~B.}\ \bibnamefont
  {Kaplan}},\ }\href {\doibase 10.1016/0920-5632(93)90282-B} {\bibfield
  {journal} {\bibinfo  {journal} {Nucl. Phys. Proc. Suppl.}\ }\textbf {\bibinfo
  {volume} {30}},\ \bibinfo {pages} {597} (\bibinfo {year} {1993})}\BibitemShut
  {NoStop}%
%%CITATION = NUPHZ,30,597;%%
\bibitem [{\citenamefont {Neuberger}(1998{\natexlab{c}})}]{Neuberger:1997bg}%
  \BibitemOpen
  \bibfield  {author} {\bibinfo {author} {\bibfnamefont {H.}~\bibnamefont
  {Neuberger}},\ }\href {\doibase 10.1103/PhysRevD.57.5417} {\bibfield
  {journal} {\bibinfo  {journal} {Phys. Rev.}\ }\textbf {\bibinfo {volume}
  {D57}},\ \bibinfo {pages} {5417} (\bibinfo {year} {1998}{\natexlab{c}})},\
  \Eprint {http://arxiv.org/abs/hep-lat/9710089} {arXiv:hep-lat/9710089
  [hep-lat]} \BibitemShut {NoStop}%
%%CITATION = HEP-LAT/9710089;%%
\bibitem [{\citenamefont {Kaplan}\ and\ \citenamefont
  {Schmaltz}(2000)}]{Kaplan:1999jn}%
  \BibitemOpen
  \bibfield  {author} {\bibinfo {author} {\bibfnamefont {D.~B.}\ \bibnamefont
  {Kaplan}}\ and\ \bibinfo {author} {\bibfnamefont {M.}~\bibnamefont
  {Schmaltz}},\ }\bibfield  {booktitle} {\emph {\bibinfo {booktitle} {{Chiral
  gauge theories. Proceedings, Workshop, Chiral'99, Taipei, Taiwan, September
  13-18, 1999}}},\ }\href@noop {} {\bibfield  {journal} {\bibinfo  {journal}
  {Chin. J. Phys.}\ }\textbf {\bibinfo {volume} {38}},\ \bibinfo {pages} {543}
  (\bibinfo {year} {2000})},\ \Eprint {http://arxiv.org/abs/hep-lat/0002030}
  {arXiv:hep-lat/0002030 [hep-lat]} \BibitemShut {NoStop}%
%%CITATION = HEP-LAT/0002030;%%
\bibitem [{\citenamefont {Kane}\ and\ \citenamefont
  {Mele}(2005)}]{kane2005quantum}%
  \BibitemOpen
  \bibfield  {author} {\bibinfo {author} {\bibfnamefont {C.~L.}\ \bibnamefont
  {Kane}}\ and\ \bibinfo {author} {\bibfnamefont {E.~J.}\ \bibnamefont
  {Mele}},\ }\href@noop {} {\bibfield  {journal} {\bibinfo  {journal} {Phys.
  Rev. Lett.}\ }\textbf {\bibinfo {volume} {95}},\ \bibinfo {pages} {226801}
  (\bibinfo {year} {2005})}\BibitemShut {NoStop}%
\bibitem [{\citenamefont {Kitaev}(2001)}]{kitaev2001unpaired}%
  \BibitemOpen
  \bibfield  {author} {\bibinfo {author} {\bibfnamefont {A.~Y.}\ \bibnamefont
  {Kitaev}},\ }\href@noop {} {\bibfield  {journal} {\bibinfo  {journal}
  {Physics-Uspekhi}\ }\textbf {\bibinfo {volume} {44}},\ \bibinfo {pages} {131}
  (\bibinfo {year} {2001})}\BibitemShut {NoStop}%
\bibitem [{\citenamefont {Atiyah}\ and\ \citenamefont
  {Bott}(1983)}]{atiyah1983yang}%
  \BibitemOpen
  \bibfield  {author} {\bibinfo {author} {\bibfnamefont {M.~F.}\ \bibnamefont
  {Atiyah}}\ and\ \bibinfo {author} {\bibfnamefont {R.}~\bibnamefont {Bott}},\
  }\href@noop {} {\bibfield  {journal} {\bibinfo  {journal} {Philosophical
  Transactions of the Royal Society of London A: Mathematical, Physical and
  Engineering Sciences}\ }\textbf {\bibinfo {volume} {308}},\ \bibinfo {pages}
  {523} (\bibinfo {year} {1983})}\BibitemShut {NoStop}%
\bibitem [{\citenamefont {Narayanan}\ and\ \citenamefont
  {Neuberger}(2006)}]{Narayanan:2006rf}%
  \BibitemOpen
  \bibfield  {author} {\bibinfo {author} {\bibfnamefont {R.}~\bibnamefont
  {Narayanan}}\ and\ \bibinfo {author} {\bibfnamefont {H.}~\bibnamefont
  {Neuberger}},\ }\href {\doibase 10.1088/1126-6708/2006/03/064} {\bibfield
  {journal} {\bibinfo  {journal} {JHEP}\ }\textbf {\bibinfo {volume} {03}},\
  \bibinfo {pages} {064} (\bibinfo {year} {2006})},\ \Eprint
  {http://arxiv.org/abs/hep-th/0601210} {arXiv:hep-th/0601210 [hep-th]}
  \BibitemShut {NoStop}%
%%CITATION = HEP-TH/0601210;%%
\bibitem [{\citenamefont {Luscher}(2010)}]{Luscher:2010iy}%
  \BibitemOpen
  \bibfield  {author} {\bibinfo {author} {\bibfnamefont {M.}~\bibnamefont
  {Luscher}},\ }\href {\doibase 10.1007/JHEP08(2010)071,
  10.1007/JHEP03(2014)092} {\bibfield  {journal} {\bibinfo  {journal} {JHEP}\
  }\textbf {\bibinfo {volume} {1008}},\ \bibinfo {pages} {071} (\bibinfo {year}
  {2010})},\ \Eprint {http://arxiv.org/abs/1006.4518} {arXiv:1006.4518
  [hep-lat]} \BibitemShut {NoStop}%
%%CITATION = ARXIV:1006.4518;%%
\bibitem [{\citenamefont {Okumura}\ and\ \citenamefont
  {Suzuki}(2016)}]{Okumura:2016dsr}%
  \BibitemOpen
  \bibfield  {author} {\bibinfo {author} {\bibfnamefont {K.-i.}\ \bibnamefont
  {Okumura}}\ and\ \bibinfo {author} {\bibfnamefont {H.}~\bibnamefont
  {Suzuki}},\ }\href@noop {} {\  (\bibinfo {year} {2016})},\ \Eprint
  {http://arxiv.org/abs/1608.02217} {arXiv:1608.02217 [hep-lat]} \BibitemShut
  {NoStop}%
%%CITATION = ARXIV:1608.02217;%%
\bibitem [{\citenamefont {Luscher}(1982)}]{luscher1982}%
  \BibitemOpen
  \bibfield  {author} {\bibinfo {author} {\bibfnamefont {M.}~\bibnamefont
  {Luscher}},\ }\href {http://projecteuclid.org/euclid.cmp/1103921338}
  {\bibfield  {journal} {\bibinfo  {journal} {Comm. Math. Phys.}\ }\textbf
  {\bibinfo {volume} {85}},\ \bibinfo {pages} {39} (\bibinfo {year}
  {1982})}\BibitemShut {NoStop}%
\bibitem [{\citenamefont {Teper}(1985)}]{Teper:1985rb}%
  \BibitemOpen
  \bibfield  {author} {\bibinfo {author} {\bibfnamefont {M.}~\bibnamefont
  {Teper}},\ }\href {\doibase 10.1016/0370-2693(85)90939-6} {\bibfield
  {journal} {\bibinfo  {journal} {Phys. Lett.}\ }\textbf {\bibinfo {volume}
  {B162}},\ \bibinfo {pages} {357} (\bibinfo {year} {1985})}\BibitemShut
  {NoStop}%
%%CITATION = PHLTA,B162,357;%%
\bibitem [{\citenamefont {Teper}(1986{\natexlab{a}})}]{Teper:1985gi}%
  \BibitemOpen
  \bibfield  {author} {\bibinfo {author} {\bibfnamefont {M.}~\bibnamefont
  {Teper}},\ }\href {\doibase 10.1016/0370-2693(86)91003-8} {\bibfield
  {journal} {\bibinfo  {journal} {Phys. Lett.}\ }\textbf {\bibinfo {volume}
  {B171}},\ \bibinfo {pages} {81} (\bibinfo {year}
  {1986}{\natexlab{a}})}\BibitemShut {NoStop}%
%%CITATION = PHLTA,B171,81;%%
\bibitem [{\citenamefont {Teper}(1986{\natexlab{b}})}]{Teper:1985ek}%
  \BibitemOpen
  \bibfield  {author} {\bibinfo {author} {\bibfnamefont {M.}~\bibnamefont
  {Teper}},\ }\href {\doibase 10.1016/0370-2693(86)91004-X} {\bibfield
  {journal} {\bibinfo  {journal} {Phys. Lett.}\ }\textbf {\bibinfo {volume}
  {B171}},\ \bibinfo {pages} {86} (\bibinfo {year}
  {1986}{\natexlab{b}})}\BibitemShut {NoStop}%
%%CITATION = PHLTA,B171,86;%%
\bibitem [{\citenamefont {Braam}\ and\ \citenamefont {van
  Baal}(1989)}]{Braam:1988qk}%
  \BibitemOpen
  \bibfield  {author} {\bibinfo {author} {\bibfnamefont {P.~J.}\ \bibnamefont
  {Braam}}\ and\ \bibinfo {author} {\bibfnamefont {P.}~\bibnamefont {van
  Baal}},\ }\href {\doibase 10.1007/BF01257416} {\bibfield  {journal} {\bibinfo
   {journal} {Commun. Math. Phys.}\ }\textbf {\bibinfo {volume} {122}},\
  \bibinfo {pages} {267} (\bibinfo {year} {1989})}\BibitemShut {NoStop}%
%%CITATION = CMPHA,122,267;%%
\bibitem [{\citenamefont {Garcia~Perez}\ \emph {et~al.}(1990)\citenamefont
  {Garcia~Perez}, \citenamefont {Gonzalez-Arroyo},\ and\ \citenamefont
  {Soderberg}}]{GarciaPerez:1989gt}%
  \BibitemOpen
  \bibfield  {author} {\bibinfo {author} {\bibfnamefont {M.}~\bibnamefont
  {Garcia~Perez}}, \bibinfo {author} {\bibfnamefont {A.}~\bibnamefont
  {Gonzalez-Arroyo}}, \ and\ \bibinfo {author} {\bibfnamefont {B.}~\bibnamefont
  {Soderberg}},\ }\href {\doibase 10.1016/0370-2693(90)90106-G} {\bibfield
  {journal} {\bibinfo  {journal} {Phys. Lett.}\ }\textbf {\bibinfo {volume}
  {B235}},\ \bibinfo {pages} {117} (\bibinfo {year} {1990})}\BibitemShut
  {NoStop}%
%%CITATION = PHLTA,B235,117;%%
\bibitem [{\citenamefont {de~Forcrand}\ \emph {et~al.}(1997)\citenamefont
  {de~Forcrand}, \citenamefont {Garcia~Perez},\ and\ \citenamefont
  {Stamatescu}}]{deForcrand:1997esx}%
  \BibitemOpen
  \bibfield  {author} {\bibinfo {author} {\bibfnamefont {P.}~\bibnamefont
  {de~Forcrand}}, \bibinfo {author} {\bibfnamefont {M.}~\bibnamefont
  {Garcia~Perez}}, \ and\ \bibinfo {author} {\bibfnamefont {I.-O.}\
  \bibnamefont {Stamatescu}},\ }\href {\doibase 10.1016/S0550-3213(97)00275-7}
  {\bibfield  {journal} {\bibinfo  {journal} {Nucl. Phys.}\ }\textbf {\bibinfo
  {volume} {B499}},\ \bibinfo {pages} {409} (\bibinfo {year} {1997})},\ \Eprint
  {http://arxiv.org/abs/hep-lat/9701012} {arXiv:hep-lat/9701012 [hep-lat]}
  \BibitemShut {NoStop}%
%%CITATION = HEP-LAT/9701012;%%
\bibitem [{\citenamefont {Garcia~Perez}\ \emph {et~al.}(1994)\citenamefont
  {Garcia~Perez}, \citenamefont {Gonzalez-Arroyo}, \citenamefont {Snippe},\
  and\ \citenamefont {van Baal}}]{GarciaPerez:1993lic}%
  \BibitemOpen
  \bibfield  {author} {\bibinfo {author} {\bibfnamefont {M.}~\bibnamefont
  {Garcia~Perez}}, \bibinfo {author} {\bibfnamefont {A.}~\bibnamefont
  {Gonzalez-Arroyo}}, \bibinfo {author} {\bibfnamefont {J.~R.}\ \bibnamefont
  {Snippe}}, \ and\ \bibinfo {author} {\bibfnamefont {P.}~\bibnamefont {van
  Baal}},\ }\href {\doibase 10.1016/0550-3213(94)90631-9} {\bibfield  {journal}
  {\bibinfo  {journal} {Nucl. Phys.}\ }\textbf {\bibinfo {volume} {B413}},\
  \bibinfo {pages} {535} (\bibinfo {year} {1994})},\ \Eprint
  {http://arxiv.org/abs/hep-lat/9309009} {arXiv:hep-lat/9309009 [hep-lat]}
  \BibitemShut {NoStop}%
%%CITATION = HEP-LAT/9309009;%%
\bibitem [{\citenamefont {Vranas}(1998)}]{Vranas:1997da}%
  \BibitemOpen
  \bibfield  {author} {\bibinfo {author} {\bibfnamefont {P.~M.}\ \bibnamefont
  {Vranas}},\ }\href {\doibase 10.1103/PhysRevD.57.1415} {\bibfield  {journal}
  {\bibinfo  {journal} {Phys. Rev.}\ }\textbf {\bibinfo {volume} {D57}},\
  \bibinfo {pages} {1415} (\bibinfo {year} {1998})},\ \Eprint
  {http://arxiv.org/abs/hep-lat/9705023} {arXiv:hep-lat/9705023 [hep-lat]}
  \BibitemShut {NoStop}%
%%CITATION = HEP-LAT/9705023;%%
\bibitem [{\citenamefont {Kikukawa}\ and\ \citenamefont
  {Noguchi}(2000)}]{Kikukawa:1999sy}%
  \BibitemOpen
  \bibfield  {author} {\bibinfo {author} {\bibfnamefont {Y.}~\bibnamefont
  {Kikukawa}}\ and\ \bibinfo {author} {\bibfnamefont {T.}~\bibnamefont
  {Noguchi}},\ }\bibfield  {booktitle} {\emph {\bibinfo {booktitle} {{Lattice
  field theory. Proceedings, 17th International Symposium, Lattice'99, Pisa,
  Italy, June 29-July 3, 1999}}},\ }\href {\doibase
  10.1016/S0920-5632(00)91758-4} {\bibfield  {journal} {\bibinfo  {journal}
  {Nucl. Phys. Proc. Suppl.}\ }\textbf {\bibinfo {volume} {83}},\ \bibinfo
  {pages} {630} (\bibinfo {year} {2000})},\ \Eprint
  {http://arxiv.org/abs/hep-lat/9902022} {arXiv:hep-lat/9902022 [hep-lat]}
  \BibitemShut {NoStop}%
%%CITATION = HEP-LAT/9902022;%%
\bibitem [{\citenamefont {Fujikawa}(1999)}]{Fujikawa:1998if}%
  \BibitemOpen
  \bibfield  {author} {\bibinfo {author} {\bibfnamefont {K.}~\bibnamefont
  {Fujikawa}},\ }\href {\doibase 10.1016/S0550-3213(99)00042-5} {\bibfield
  {journal} {\bibinfo  {journal} {Nucl. Phys.}\ }\textbf {\bibinfo {volume}
  {B546}},\ \bibinfo {pages} {480} (\bibinfo {year} {1999})},\ \Eprint
  {http://arxiv.org/abs/hep-th/9811235} {arXiv:hep-th/9811235 [hep-th]}
  \BibitemShut {NoStop}%
%%CITATION = HEP-TH/9811235;%%
\bibitem [{\citenamefont {Adams}(2002)}]{Adams:1998eg}%
  \BibitemOpen
  \bibfield  {author} {\bibinfo {author} {\bibfnamefont {D.~H.}\ \bibnamefont
  {Adams}},\ }\href {\doibase 10.1006/aphy.2001.6209} {\bibfield  {journal}
  {\bibinfo  {journal} {Annals Phys.}\ }\textbf {\bibinfo {volume} {296}},\
  \bibinfo {pages} {131} (\bibinfo {year} {2002})},\ \Eprint
  {http://arxiv.org/abs/hep-lat/9812003} {arXiv:hep-lat/9812003 [hep-lat]}
  \BibitemShut {NoStop}%
%%CITATION = HEP-LAT/9812003;%%
\bibitem [{\citenamefont {Suzuki}(1999)}]{Suzuki:1998yz}%
  \BibitemOpen
  \bibfield  {author} {\bibinfo {author} {\bibfnamefont {H.}~\bibnamefont
  {Suzuki}},\ }\href {\doibase 10.1143/PTP.102.141} {\bibfield  {journal}
  {\bibinfo  {journal} {Prog. Theor. Phys.}\ }\textbf {\bibinfo {volume}
  {102}},\ \bibinfo {pages} {141} (\bibinfo {year} {1999})},\ \Eprint
  {http://arxiv.org/abs/hep-th/9812019} {arXiv:hep-th/9812019 [hep-th]}
  \BibitemShut {NoStop}%
%%CITATION = HEP-TH/9812019;%%
\bibitem [{\citenamefont {Makino}\ and\ \citenamefont
  {Morikawa}(2016)}]{Makino:2016auf}%
  \BibitemOpen
  \bibfield  {author} {\bibinfo {author} {\bibfnamefont {H.}~\bibnamefont
  {Makino}}\ and\ \bibinfo {author} {\bibfnamefont {O.}~\bibnamefont
  {Morikawa}},\ }\href@noop {} {\  (\bibinfo {year} {2016})},\ \Eprint
  {http://arxiv.org/abs/1609.08376} {arXiv:1609.08376 [hep-lat]} \BibitemShut
  {NoStop}%
%%CITATION = ARXIV:1609.08376;%%
\bibitem [{\citenamefont {Fukaya}\ \emph {et~al.}(2016)\citenamefont {Fukaya},
  \citenamefont {Onogi}, \citenamefont {Yamamoto},\ and\ \citenamefont
  {Yamamura}}]{Fukaya:2016ofi}%
  \BibitemOpen
  \bibfield  {author} {\bibinfo {author} {\bibfnamefont {H.}~\bibnamefont
  {Fukaya}}, \bibinfo {author} {\bibfnamefont {T.}~\bibnamefont {Onogi}},
  \bibinfo {author} {\bibfnamefont {S.}~\bibnamefont {Yamamoto}}, \ and\
  \bibinfo {author} {\bibfnamefont {R.}~\bibnamefont {Yamamura}},\ }\href@noop
  {} {\  (\bibinfo {year} {2016})},\ \Eprint {http://arxiv.org/abs/1607.06174}
  {arXiv:1607.06174 [hep-th]} \BibitemShut {NoStop}%
%%CITATION = ARXIV:1607.06174;%%
\bibitem [{\citenamefont {Luscher}(2000)}]{Luscher:1999un}%
  \BibitemOpen
  \bibfield  {author} {\bibinfo {author} {\bibfnamefont {M.}~\bibnamefont
  {Luscher}},\ }\href {\doibase 10.1016/S0550-3213(99)00731-2} {\bibfield
  {journal} {\bibinfo  {journal} {Nucl. Phys.}\ }\textbf {\bibinfo {volume}
  {B568}},\ \bibinfo {pages} {162} (\bibinfo {year} {2000})},\ \Eprint
  {http://arxiv.org/abs/hep-lat/9904009} {arXiv:hep-lat/9904009 [hep-lat]}
  \BibitemShut {NoStop}%
%%CITATION = HEP-LAT/9904009;%%
\bibitem [{\citenamefont {Luscher}(1999)}]{Luscher:1998du}%
  \BibitemOpen
  \bibfield  {author} {\bibinfo {author} {\bibfnamefont {M.}~\bibnamefont
  {Luscher}},\ }\href {\doibase 10.1016/S0550-3213(99)00115-7} {\bibfield
  {journal} {\bibinfo  {journal} {Nucl. Phys.}\ }\textbf {\bibinfo {volume}
  {B549}},\ \bibinfo {pages} {295} (\bibinfo {year} {1999})},\ \Eprint
  {http://arxiv.org/abs/hep-lat/9811032} {arXiv:hep-lat/9811032 [hep-lat]}
  \BibitemShut {NoStop}%
%%CITATION = HEP-LAT/9811032;%%
\end{thebibliography}%
\end{document}